\documentclass[lettersize,journal]{IEEEtran}

\usepackage{amsmath, amsfonts}

\usepackage{algorithmic}
\usepackage{algorithm}

\usepackage{array}
\usepackage{booktabs}
\usepackage{multirow}

\usepackage{graphicx}
\graphicspath{{Figures/}}

\usepackage{float}
\usepackage[caption=false, font=normalsize, labelfont=sf, textfont=sf]{subfig} 
\usepackage[font=small, labelfont=it]{caption} 
\usepackage{adjustbox} 

\providecommand{\columnwidth}{0.32\textwidth}

\usepackage{tikz}
\usetikzlibrary{arrows.meta, shapes.geometric, calc, positioning}

\usepackage{textcomp}
\usepackage{stfloats}
\usepackage{url}
\usepackage{verbatim}

\usepackage{cite}

\usepackage{siunitx}

\begin{document}

\title{Partial-Power Flow Controller, Voltage Regulator, and Energy Router for Hybrid AC--DC Grids}

\author{
    \IEEEauthorblockN{Ehsan Asadi, Davood Keshavarzi, Alexander Koehler, Nima Tashakor, and Stefan Goetz}
}
\maketitle
\begin{abstract}
The share of electronically converted power from renewable sources, loads, and storage is continuously growing in the low- and medium-voltage grids. These sources and loads typically rectify the grid AC to DC, e.g., for a DC link, so that a DC grid could eliminate hardware and losses of these conversion stages. However, extended DC grids lack the stabilizing nature of AC impedances so that the voltage is more fragile and power flows may need active control, particularly if redundancy as known from AC, such as rings and meshing, is desired. Furthermore, a DC infrastructure will not replace but will need to interface with the existing AC grid.

This paper presents a partial-power energy router architecture that can interface multiple AC and DC lines to enable precise control of voltages and both active as well as reactive power flows. The proposed system uses modular low-voltage high-current series modules supplied through dual active bridges. These modules only need to process a small share of the voltage to control large power flows. The topology reduces component size, cost, energy losses, and reliability more than three times compared to conventional technology. The optional integration of battery energy storage can furthermore eliminate the need for the sum of the power flows of all inputs to be zero at all times. Through dynamic voltage injection relative to the line voltage, the modules effectively balance feeder currents, regulate reactive power, and improve the power factor in AC grids.

Real-time hardware-in-the-loop and prototype measurements validate the proposed energy router's performance under diverse operating conditions. Experimental results confirm the series module's functionality in both AC and DC grids as an effective solution for controlling extended grids, including power sharing, voltage, and power quality.
\end{abstract}


\section{Introduction}
Modern power systems would benefit from meshed microgrids, which connect multiple feeders and distributed energy resources to boost redundancy, flexibility, as well as reliability and manage the rapidly increased share of distributed generation. 
However, these trends create a range of critical technical challenges. In AC grids, meshed configurations introduce complex bidirectional power flows, which compromise voltage stability and increase the risk of line overload, especially when they interact with intermittent sources and high-power loads \cite{SGoetz,  richardson2010electric, sepehrzad2023}. Power sharing in these networks remains inaccurate due to mismatched line impedances, voltage deviations, and converter nonidealities. These issues cause circulating currents, uneven load distribution, and degraded operational efficiency \cite{benhassi2024, abouassi2023, sepehrzad2023}.
Conventional droop control, while decentralized and simple, suffers from several drawbacks: it leads to voltage and frequency deviations, poor transient response, and circulating currents. These limitations make it unsuitable for ensuring stable and efficient operation in meshed low-voltage networks \cite{mishra2024, rashwan2023, benhassi2024, Chandak2022, Guerrero2011, payghami, mokhtari, Zhao}.

DC grids, though attractive for last-mile efficiency and renewable integration, face their own challenges.
 DC grids lack lossless impedance for stabilization. 
Whereas inductances play that role in AC grids, inductances  only act transiently in DC and can worsen fluctuations. The consequence is an increased sensitivity to voltage variations and ripple effects from converters and loads. Moreover, stability issues arise when DC grids interact with other DC or AC grids; such interactions often result in oscillations and poor damping \cite{deng2021interaction}.

Furthermore, AC--DC grid interfaces can conflict with protection paradigms. Whereas AC grids need a minimum fault current for protection, such as fuses, to react, the growing consensus for DC grids suggests detection of short circuits and immediate reduction of current fed into it. Consequently, the uncontrolled capacitance in the line should be minimum for effective fault protection and system stability \cite{SGoetz, Lu2024}.

Finally, constant-power loads (CPLs) such as EV chargers and data center equipment contribute negative incremental impedance, leading to instability and power oscillations \cite{blaabjerg2006overview, Kwasinski2011, Middlebrook1976, kacetldc,Kwasinski2011b,electric_springs,deng2021interaction,tu2023impact}.

Conventional solutions for the stabilization of the grid voltage, such as on-load tap changers and shunt capacitor banks, are too slow,  lack the control bandwidth required to solve these issues effectively, and do not work in DC. Furthermore, such solutions can conflict with protection paradigms. Whereas AC grids need a minimum fault current for protection, such as fuses, to react, the growing consensus for DC grids suggests detection of short circuits and immediate reduction of current fed into it. Consequently, the uncontrolled capacitance in the line should be minimal for effective fault protection and system stability.
 
Active filters and series voltage compensators can mitigate ripple to maintain system efficiency, protection, and stability \cite{SeriesVoltageCompensator,UltraCompactPowerBuffer}.
 Electric springs dynamically adjust impedance to stabilize voltage and improve load behavior \cite{electric_springs}, and SOPs enable flexible power flow between distributed feeders, and voltage stability \cite{overview_SOP, delta_SNOP, series_shunt_SNOP}. These advancements mark a shift toward active, dynamic control strategies that solve the problems of modern grids.
 
Alternative control methods at the connected loads and sources cannot solve the problems. Instead, multiport and meshed DC grids require control of the power flow. In AC transmission lines, 
unified power-flow controllers (UPFCs) can support or even counter passive distribution of power flows according to the complex impedances \cite{lu2016optimal, chen2013transformerless, guan2015transformerless}. They combine series and shunt compensation to regulate voltage and phase across high-voltage transmission networks \cite{gyugyi1992unified, hingorani2000understanding, zhang2012flexible}. 
So-called soft open points can provide similar control on the AC distribution level, but typically are entire AC--DC--AC converters that need to process the full power, i.e., full voltage and full current \cite{peng, Zhang}. Lu et al., however, proposed a universal direct-injection circuit with partial-power capability to regulate both power flow and power quality with only very small voltage injections and a fraction of the power \cite{Lu2023,Lu2024,Lu2024b, Lu2023b}. 

For DC grids, power-flow control is practically unknown. Almost all reports in the literature assume either single-point DC systems or a line, where all nodes have same or similar voltage and power-flow imbalances are not possible. Few reports extend the grid and then typically include full-power conversion \cite{Liang2019}.  Su and Li subsequently derived  a differential voltage processing framework to equalize voltage and current in multi-leg battery configurations \cite{DPP}. The system still assumes a simple DC line, but translates the concept of series injection to DC. Similarly, some vehicle DC charging concepts fine-tune the vehicle-side output voltage voltage with differential concepts \cite{Rivera2022, RiveraCharging}. However, a compact power-flow hub and interface between multiple AC and DC lines is still missing.

\begin{figure*}[t]
    \centering

\includegraphics[width=2.0\columnwidth]{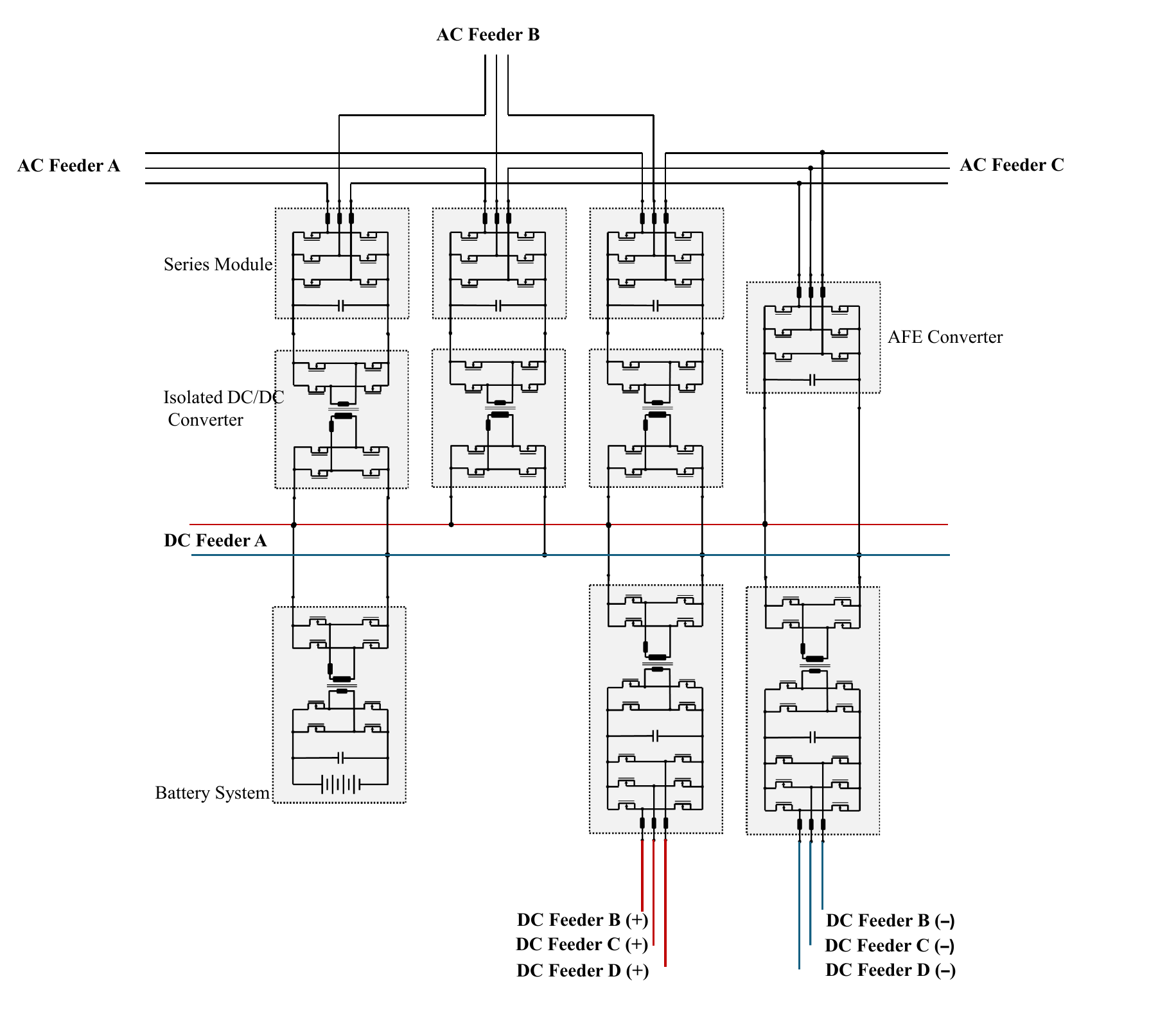}

    \caption{Multiline multi-function AC--DC energy router.}
    \label{fig:Energy Hub2}
\end{figure*}

This paper presents a novel energy router architecture (Fig.~\ref{fig:Energy Hub2}) designed for hybrid AC--DC grids, with a low-voltage multi-leg series module topology to enable bidirectional power flow, integrate renewable energy sources, and balance dynamic loads. The design corrects phase and voltage mismatches, reduces ripple voltage, and can use compact, high-reliability film capacitors to achieve lower short-circuit currents, while it maintains voltage stability. Integrated current control, voltage regulation, and impedance adaptation provide virtual inertia, transient damping, ripple mitigation, and fault recovery. This design solves the key problems of modern hybrid grids and enables scalable, reliable operation under evolving power system demands. 
The paper is structured as follows: Section II presents the hybrid network architecture and multi-leg topology. Sections III–V describe the proposed control strategies. Sections VI–VIII evaluate the system’s performance and show improvements in power sharing and stability. Sections IX and X present case studies and experimental validation.

\section{Energy Hub Architecture and Power Flow Control}
The energy router can connect multiple AC and DC feeders to distribute power, regulate active and reactive power independently, and stabilize voltages. Using partial power processing (PPP), series modules inject a small voltage to control large power flows due to low line impedances, even in weak grids. It is scalable and requires only additional half bridges for new feeders. Power transfer from Feeder \( i \) to Feeder \( j \) is \( P_i + P_{\text{series},k} \), with \( P_{\text{series},k} \ll P_i \), defined as
\begin{align}
P_{\text{transfer},k} &= V_{\text{bus}} \cdot I_{\text{line},k}, \label{eq:p_transfer} \\
P_{\text{series},k} &= V_{\text{inj},k} \cdot I_{\text{line},k}, \label{eq:p_converter}
\end{align}
where \( V_{\text{inj},k} \ll V_{\text{bus}} \). The suggested partial power processing solution reduces the necessary component ratings, stress, and losses compared to full-power DAB or buck--boost designs. Consequently, partial power processing enables a compact, cost-effective system. A 200 kW transmission system comparison, based on MIL-HDBK-217F, IEC 61709, and IEC 62380 standards \cite{MIL217, IEC 61709, IEC 62380}, demonstrates a 43\% cost reduction and 3.1-fold reliability improvement over conventional systems. Table~\ref{tab:aspect-comparison} summarizes efficiency, scalability, and cost comparison of the other alternatives compared to the partial power processing.
Figure~\ref{fig:Energy Hub2} depicts the multiline multi-function AC--DC energy router, with a three-feeder AC and three-feeder DC configuration.
\subsection{Key Components}
The router integrates the following core components:
\begin{itemize}
    \item \textbf{Active Front-End Converter (AFE):} A shunt inverter that maintains the DC-link voltage and enables bidirectional AC-DC power exchange \cite{Khan2023, Parker2014}. It powers Dual Active Bridges (DABs) and series modules while regulating reactive power (see Fig.~\ref{fig:shunt_converter_controller}).
    \begin{figure}[h]
        \centering
        \includegraphics[width=\columnwidth]{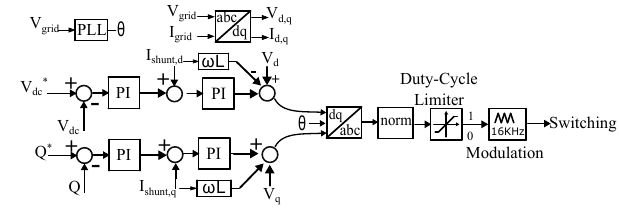}
        \caption{Shunt active front-end converter control.}
        \label{fig:shunt_converter_controller}
    \end{figure}
    \item \textbf{Dual Active Bridge Converters:} These provide galvanic isolation and supply floating voltages to the series modules \cite{Muhlethaler2013, Wang2016}.
    \item \textbf{Series Modules:} These generate PWM-regulated AC/DC voltages to track power references, compensate for voltage and phase mismatches, and dampen oscillations, utilizing only a fraction of the grid voltage.
\end{itemize}
\begin{table}[t]
    \centering
    \caption{Power Processing Comparison}
    \label{tab:aspect-comparison}
    \adjustbox{width=\columnwidth}{
    \begin{tabular}{@{}lccc@{}}
        \toprule
        \textbf{Aspect} & \textbf{PPP} & \textbf{DAB (Full)} & \textbf{Buck--Boost (Full)} \\
        \midrule
        Power Processed & \SI{10}{\percent} & \SI{100}{\percent} & \SI{100}{\percent} \\
        Voltage Stress & Low & High & Line Voltage \\
        Efficiency & \textgreater \SI{99}{\percent} & \SI{95}{\percent} & \SI{96}{\percent}--\SI{98}{\percent} \\
        Bidirectional & Yes & Yes & Limited \\
        Isolation & Yes & Yes & No \\
        Component Size & Small & Large & Compact \\
        Cost & Low & High & Moderate \\
        \bottomrule
    \end{tabular}}
\end{table}

\section{Multi-Feeder AC Grid Analysis}

A multi-feeder AC grid with \( N \) feeders is analyzed, where Feeder 1 powers an AFE rectifier. Each feeder \( k \) (\( k = 1, 2, \dots, N \)) connects to a series module injecting voltage \( V_{\text{inj},k} \angle \phi_{\text{inj},k} \). Dual Active Bridge (DAB) converters support the DC link for all series modules, drawing power from the AFE rectifier, which converts AC from Feeder 1 to a DC bus voltage \( V_{dc} \). An Energy Storage System (ESS) buffers transient power variations.

\subsection{Power Flow Model}

For each feeder \( k \), power flow between the feeder and the AFE’s AC terminal (bus) is given by:
\begin{align}
P_k &= \text{Re} \left[ V_k I_{\text{line},k}^* \right], \label{eq:p_k} \\
Q_k &= \text{Im} \left[ V_k I_{\text{line},k}^* \right], \label{eq:q_k}
\end{align}
where the line current is:
\begin{equation}
I_{\text{line},k} = \frac{V_k - V_{\text{bus,AC}} - V_{\text{inj},k}}{Z_k},
\end{equation}
with \( V_k \angle \delta_k \) as the feeder voltage, \( V_{\text{bus,AC}} \angle \delta_{\text{bus}} \) as the AFE’s AC terminal voltage, \( V_{\text{inj},k} \angle \phi_{\text{inj},k} \) as the series-injected voltage, and \( Z_k = R_k + j X_k \) as the line impedance. The line current magnitude is:
\begin{equation}
|I_{\text{line},k}| = \frac{|V_k - V_{\text{bus,AC}} - V_{\text{inj},k}|}{|Z_k|},
\end{equation}
and its phase \( \theta_{I,k} \) depends on \( \angle (V_k - V_{\text{bus,AC}} - V_{\text{inj},k}) - \angle Z_k \).

Using phasor analysis and including resistance, the active and reactive power are:
\begin{align}
P_k &= \frac{|V_k| |V_{\text{bus,AC}}|}{|Z_k|} \left[ \cos(\angle Z_k) \cos(\delta_k - \delta_{\text{bus}}) \right. \notag \\
&\quad \left. + \sin(\angle Z_k) \sin(\delta_k - \delta_{\text{bus}}) \right] \notag \\
&\quad + |V_{\text{inj},k}| \cdot |I_{\text{line},k}| \cos(\phi_{\text{inj},k} - \theta_{I,k}), \label{eq:p_k_full} \\
Q_k &= \frac{|V_k| |V_{\text{bus,AC}}|}{|Z_k|} \left[ \sin(\angle Z_k) \cos(\delta_k - \delta_{\text{bus}}) \right. \notag \\
&\quad \left. - \cos(\angle Z_k) \sin(\delta_k - \delta_{\text{bus}}) \right] \notag \\
&\quad + |V_{\text{inj},k}| \cdot |I_{\text{line},k}| \sin(\phi_{\text{inj},k} - \theta_{I,k}), \label{eq:q_k_full}
\end{align}
where \( \angle Z_k = \tan^{-1}(X_k / R_k) \).

\subsection{dq-Frame Control}

The router employs a dq-reference frame, aligning with \( V_{\text{bus,AC}} \) (\( \delta_{\text{bus}} = 0 \)):
\begin{equation}
V_{\text{inj},k} \angle \phi_{\text{inj},k} = V_{\text{inj},d,k} + j V_{\text{inj},q,k}.
\end{equation}
For small phase differences (\( \delta_k - \delta_{\text{bus}} \approx 0 \)) and general impedance, the power expressions simplify to:
\begin{align}
P_k &\approx \frac{|V_k| |V_{\text{bus,AC}}|}{|Z_k|} \left[ \cos(\angle Z_k) + \sin(\angle Z_k) \delta_k \right] \notag \\
&\quad + \frac{|V_k| V_{\text{inj},d,k}}{|Z_k|} \cos(\angle Z_k) + \Delta P_{\text{cross}}, \label{eq:p_k_approx} \\
Q_k &\approx \frac{|V_k| (|V_k| - |V_{\text{bus,AC}}|)}{|Z_k|} \sin(\angle Z_k) \notag \\
&\quad + \frac{|V_k| V_{\text{inj},q,k}}{|Z_k|} \sin(\angle Z_k) + \Delta Q_{\text{cross}}, \label{eq:q_k_approx}
\end{align}
with cross-coupling terms:
\begin{align}
\Delta P_{\text{cross}} &= -\frac{|V_k| V_{\text{inj},q,k}}{|Z_k|} \cos(\angle Z_k) \notag \\
&\quad + \frac{|V_k| V_{\text{inj},d,k}}{|Z_k|} \sin(\angle Z_k) \delta_k \notag \\
&\quad - \frac{|V_k| V_{\text{inj},q,k}}{|Z_k|} \cos(\angle Z_k) \delta_k, \label{eq:delta_p} \\
\Delta Q_{\text{cross}} &= \frac{|V_k| V_{\text{inj},d,k}}{|Z_k|} \sin(\angle Z_k) \notag \\
&\quad + \frac{|V_k| V_{\text{inj},q,k}}{|Z_k|} \cos(\angle Z_k) \delta_k \notag \\
&\quad - \frac{|V_k| V_{\text{inj},d,k}}{|Z_k|} \sin(\angle Z_k) \delta_k. \label{eq:delta_q}
\end{align}
For dominant reactance (\( \angle Z_k \approx \pi/2 \), \( \cos(\angle Z_k) \approx 0 \), \( \sin(\angle Z_k) \approx 1 \)):
\begin{equation}
\Delta P_{\text{cross}} \approx 0, \quad \Delta Q_{\text{cross}} \approx \frac{|V_k| V_{\text{inj},d,k}}{|Z_k|}. \label{eq:delta_simplified}
\end{equation}

\subsection{Series Module Control}

The series module uses a half-bridge voltage-source converter with PI control to compensate for cross-coupling:
\begin{align}
v_{\text{inj},d,k}(t) &= K_p \left( i_{\text{ref},d,k}(t) - i_{d,k}(t) \right) \notag \\
&\quad + K_i \int \left( i_{\text{ref},d,k}(\tau) - i_{d,k}(\tau) \right) \mathrm{d}\tau \notag \\
&\quad + v_{m,d,k}, \label{eq:vd_control} \\
v_{\text{inj},q,k}(t) &= K_p \left( i_{\text{ref},q,k}(t) - i_{q,k}(t) \right) \notag \\
&\quad + K_i \int \left( i_{\text{ref},q,k}(\tau) - i_{q,k}(\tau) \right) \mathrm{d}\tau \notag \\
&\quad + v_{m,q,k}, \label{eq:vq_control}
\end{align}
where \( e_{d,k}(t) = i_{\text{ref},d,k}(t) - i_{d,k}(t) \), \( e_{q,k}(t) = i_{\text{ref},q,k}(t) - i_{q,k}(t) \), reference currents are \( i_{\text{ref},d,k} = P_{\text{ref},k} / |V_k| \), \( i_{\text{ref},q,k} = Q_{\text{ref},k} / |V_k| \), and gains are \( K_p = 100 \), \( K_i = 50 \) \cite{Ciobotaru2006}. Mismatch corrections include:
\begin{itemize}
    \item Phase: \( v_{m,d,k} = 2 |V_k| \sin(\Delta \theta_k / 2) \cos(\delta_k) \),
    \item Magnitude: \( v_{m,q,k} = |V_k - V_{\text{bus,AC}}| + 2 |V_k| \sin(\Delta \theta_k / 2) \sin(\delta_k) \).
\end{itemize}

\subsection{Power Regulation}

Orthogonality of d- and q-axes enables independent control, with partial derivatives:
\begin{align}
\frac{\partial P_k}{\partial V_{\text{inj},d,k}} &\approx \frac{|V_k|}{|Z_k|} \cos(\angle Z_k), \notag \\
\frac{\partial P_k}{\partial V_{\text{inj},q,k}} &\approx -\frac{|V_k|}{|Z_k|} \sin(\angle Z_k), \label{eq:partial_p} \\
\frac{\partial Q_k}{\partial V_{\text{inj},d,k}} &\approx \frac{|V_k|}{|Z_k|} \sin(\angle Z_k), \notag \\
\frac{\partial Q_k}{\partial V_{\text{inj},q,k}} &\approx \frac{|V_k|}{|Z_k|} \cos(\angle Z_k). \label{eq:partial_q}
\end{align}
These confirm decoupled regulation of \( P_k \) and \( Q_k \), with cross-coupling mitigated by dq-frame control, enabling effective power flow management across the grid.

\subsection{Power Flow Between AC and DC Grids}

The energy router enables bidirectional power flow between AC and DC feeders through the AFE converter. The AFE connects to AC feeder 1 on the AC side and to the common DC bus on the DC side. The battery energy storage system (BESS) and DC feeders connect to this DC bus via series modules for flexible power exchange, transient buffering, and system balance.

The AFE exchanges power on the AC side under dq-reference frame control. Active power \(P_{\text{AFE}}\) and reactive power \(Q_{\text{AFE}}\) equal
\begin{align}
P_{\text{AFE}} &= \frac{3}{2} (v_d i_d + v_q i_q), \label{eq:p_afe} \\
Q_{\text{AFE}} &= \frac{3}{2} (v_d i_q - v_q i_d), \label{eq:q_afe}
\end{align}
where \(v_d, v_q\) and \(i_d, i_q\) denote d- and q-axis voltages and currents.

With the reference frame aligned to grid voltage (\(v_q = 0\)), active power simplifies to
\begin{equation}
P_{\text{AFE}} = \frac{3}{2} v_d i_d. \label{eq:p_afe_simplified}
\end{equation}

On the DC side, power from the AFE equals \(P_{\text{AFE,DC}} = V_{\text{dc}} I_{\text{AFE,dc}}\), where \(V_{\text{dc}}\) denotes DC bus voltage and \(I_{\text{AFE,dc}}\) denotes DC current from the AFE. Negligible losses yield \(P_{\text{AFE}} \approx P_{\text{AFE,DC}}\).

For DC feeder \(k\), power flow equals \(P_{\text{dc},k} = V_{\text{dc}} I_{\text{line},k} + v_{\text{inj},k} I_{\text{line},k}\).

DC bus power balance, without losses, follows \(P_{\text{AFE,DC}} + P_{\text{BESS}} = \sum_k P_{\text{dc},k}\), where \(P_{\text{BESS}}\) denotes power to/from the battery (positive for discharge).

The sum of powers from AC feeders balances the AFE DC-side power plus battery contribution. Power transfer occurs between any AC feeder \(m\) and DC feeder \(n\) via series modules on both sides and AFE conversion. Series modules regulate individual feeder flows; the AFE manages AC-DC conversion.

This configuration supports AC feeders that supply DC loads (e.g., EV chargers) or DC sources (e.g., photovoltaics) that support AC grids.

\section{Modified Control of Series Modules in Multi-Line DC Grids}
\label{sec:control_strategy}

Multi-line DC distribution grids promise higher system-level efficiency, particularly in industrial applications, larger renewable energy systems, and microgrids. However, they encounter stability issues from constant power loads (CPLs) and the lack of mechanical inertia in power-electronic interfaces, which would stabilize the voltage in case of sudden power fluctuations. This section introduces an advanced control strategy for series modules to provide precise current regulation and robust stability across diverse grid conditions. We combine virtual inertia and ripple mitigation to boost performance and minimize the need for large passive components.

The injected voltage \( v_{\text{inj}} \) of the series module follows
\begin{equation}
\begin{aligned}
v_{\text{inj}} &= \underbrace{K_\textrm{p} e(t) + K_\textrm{i} \int e(t) \, \mathrm{d}t}_{\text{Power-Flow Controller}} \\
&\quad - \underbrace{K_\textrm{r} v_{\text{ripple,meas}}}_{\text{Ripple mitigation}} \\
&\quad + \underbrace{K_\textrm{C} \frac{\mathrm{d} V_{\text{dc}}}{\mathrm{d}t} + K_\textrm{L} \frac{\mathrm{d} e(t)}{\mathrm{d}t}}_{\text{Virtual inertia}} + \underbrace{v_{\text{mismatch}}}_{\text{Voltage mismatch compensation}},
\end{aligned}
\label{eq:inj_voltage}
\end{equation}
where \( e(t) = i_{\text{ref}} - i_{\text{meas}} \) is the current error between the reference \( i_{\text{ref}} \) and measured \( i_{\text{meas}} \) currents, \( v_{\text{ripple,meas}} \) is the measured ripple voltage, and \( v_{\text{mismatch}} = |V_i - V_j| \) is the feedforward term for DC voltage mismatch compensation between nodes \( i \) and \( j \), ensuring precise power flow control. This injected voltage comprises three components: a PI-based power flow controller for steady-state current regulation, ripple mitigation for voltage stability, and virtual inertia to dampen voltage transients.

\subsection{Virtual Inertia Control}
\label{subsec:virtual_inertia}
DC grids lack mechanical inertia such that power transients cause voltage fluctuations. The virtual inertia control (VIC) in Equation~\ref{eq:inj_voltage} uses
\begin{equation}
v_{\text{inj,VIC}} = K_\textrm{C} \frac{\mathrm{d} V_{\text{dc}}}{\mathrm{d}t} + K_\textrm{L} \frac{\mathrm{d} e(t)}{\mathrm{d}t},
\label{eq:vic}
\end{equation}
where \( K_\textrm{C} \) acts as a virtual capacitor to stabilize \( V_{\text{dc}} \), and \( K_\textrm{L} \) functions as an inductor to smooth the current dynamics.

\subsection{Ripple Mitigation and Energy Storage}
\label{subsec:ripple_mitigation}
The term \( -K_\textrm{r} v_{\text{ripple,meas}} \) in Equation~\ref{eq:inj_voltage} counters voltage ripples from pulsations and harmonics. The method uses a filter to isolate ripple frequencies for feed-forward compensation. For a ripple current \( \Delta I_{\text{ripple}} \) at frequency \( \omega \), the series module diminishes the ripple voltage
\begin{equation}
\Delta V_{\text{ripple}} = \frac{\Delta I_{\text{ripple}}}{\omega C_{\text{dc}}}
\label{eq:ripple}
\end{equation}
by \( v_{\text{series}} \) to \( \Delta V_{\text{effective}} = \Delta V_{\text{ripple}} - v_{\text{series}} \). The control reduces the required capacitance \( C_{\text{dc}} \) to \( C_{\text{dc,with}} \propto \Delta I_{\text{ripple}} / (\omega \Delta V_{\text{effective}}) \).
The minimum capacitance reflects and supports the different failure and protection approach in DC. Whereas in AC a minimum fault current is necessary to blow fuses and avoid stable failure currents below the tripping threshold. The inductance of the grid and transformers avoids excessive short-circuit currents though. In DC, the line inductance cannot permanently limit fault currents and the electronic interfaces can typically not tolerate even short current surges so that faults should be suppressed and stopped as soon as detected.
A battery can extend the hold-up time
\begin{equation}
t_{\text{total}} = t_{\text{capacitor}} + t_{\text{battery}},
\label{eq:holdup}
\end{equation}
with \( t_{\text{capacitor}} = 2 C_{\text{dc}} (V_{\text{dc,initial}}^2 - V_{\text{dc,min}}^2) / P_{\text{load}} \) and \( t_{\text{battery}} = Q_{\text{battery}} V_{\text{battery}} / P_{\text{load}} \), where \( t_{\text{battery}} \gg t_{\text{capacitor}} \).
  
\subsection{Stability Analysis}
\label{subsec:stability}
\begin{figure}[t]
    \centering
    \begin{minipage}[t]{0.98\columnwidth}
        \centering
        \includegraphics[width=\linewidth]{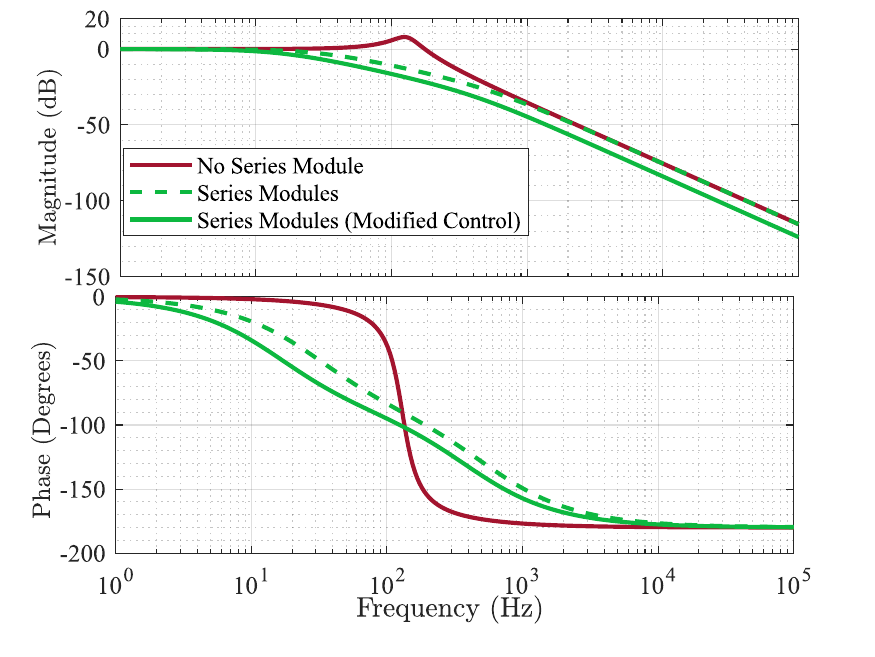}
        \caption{Bode diagram of DC-link voltage transfer functions: baseline, ripple-mitigated, and with virtual inertia.}
        \label{fig:bode1}
    \end{minipage}\hfill
    \begin{minipage}[t]{0.98\columnwidth}
        \centering
        \includegraphics[width=\linewidth]{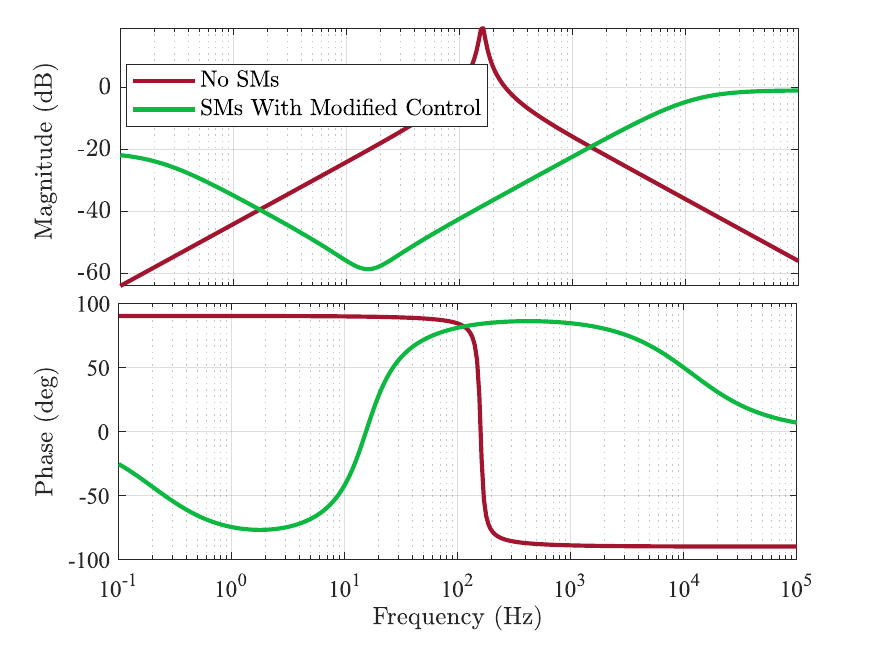}
        \caption{Bode diagram comparing DC-link voltage under normal operation and CPL connection.}
        \label{fig:bode2}
    \end{minipage}
    \vspace{-\baselineskip} 
\end{figure}

Stability of Equation~\ref{eq:inj_voltage} is evaluated via small-signal analysis. The DC-link and current dynamics are
\begin{equation}
C \frac{\mathrm{d} V_{\text{dc}}}{\mathrm{d}t} = I_{\text{dc}} - i_{\text{meas}},
\label{eq:dc_link}
\end{equation}
\begin{equation}
L \frac{\mathrm{d} i_{\text{meas}}}{\mathrm{d}t} + R i_{\text{meas}} = V_{\text{dc}} - v_{\text{inj}}.
\label{eq:current_dyn}
\end{equation}
Linearization with perturbations (\( V_{\text{dc}} = V_{\text{dc0}} + \tilde{V}_{\text{dc}} \), \( i_{\text{meas}} = i_0 + \tilde{i} \)) gives
\begin{equation}
C s \tilde{V}_{\text{dc}}(s) = -\tilde{i}(s),
\label{eq:dc_link_ss}
\end{equation}
\begin{equation}
(L s + R) \tilde{i}(s) = \tilde{V}_{\text{dc}}(s) - \tilde{v}_{\text{inj}}(s).
\label{eq:current_ss}
\end{equation}
The perturbed \( \tilde{v}_{\text{inj}}(s) \) is
\begin{equation}
\tilde{v}_{\text{inj}}(s) = G_c(s) (\tilde{i}_{\text{ref}} - \tilde{i}) + (K_\textrm{C} - K_\textrm{r}) s \tilde{V}_{\text{dc}},
\label{eq:inj_ss}
\end{equation}
where \( G_c(s) = K_\textrm{p} + K_\textrm{i}/s + K_\textrm{L} s \), and \( \tilde{v}_{\text{ripple,meas}} = s \tilde{V}_{\text{dc}} \). The substitution of this expression into Equations~\ref{eq:dc_link_ss} and \ref{eq:current_ss} yields the closed-loop transfer function as
\begin{equation}
\frac{\tilde{i}(s)}{\tilde{i}_{\text{ref}}(s)} = \frac{G_c(s)}{L s + R + \frac{1}{C s} + G_c(s) - \frac{K_\textrm{C} - K_\textrm{r}}{C}},
\label{eq:tf}
\end{equation}
with the characteristic equation
\begin{equation}
L s^2 + \left[ R + K_\textrm{p} + K_\textrm{L} - \frac{K_\textrm{C} - K_\textrm{r}}{C} \right] s + K_\textrm{i} + \frac{1}{C} = 0.
\label{eq:char}
\end{equation}
Stability requires
\begin{equation}
K_\textrm{C} - K_\textrm{r} < (R + K_\textrm{p} + K_\textrm{L}) C,
\label{eq:stability_cond}
\end{equation}
which introduces positive damping. For virtual inertia, the voltage loop with impedance \( Z \) yields
\begin{equation}
H_{\text{VIC}}(s) = -\frac{1}{\left( C + \frac{K_\textrm{C}}{Z} \right) s}.
\label{eq:vic_tf}
\end{equation}
The system is stable if \( K_\textrm{C} < Z C \). The current loop pole \( s = -R / (L + K_\textrm{L}) \) remains stable for \( R, L, K_\textrm{L} > 0 \). Bode diagrams (Figs.~\ref{fig:bode1}, \ref{fig:bode2}) confirm these findings, comparing baseline, ripple-mitigated, and inertia-enhanced responses for constant power loads.

\begin{table*}[t]
    \centering
    \caption{Comparative Analysis of the Proposed Energy Hub and Existing Solutions}
    \label{tab:comparison}
    \resizebox{\textwidth}{!}{%
    \begin{tabular}{@{}>{\raggedright\arraybackslash}p{3.2cm}%
                        >{\raggedright\arraybackslash}p{3.2cm}%
                        >{\raggedright\arraybackslash}p{2.8cm}%
                        >{\raggedright\arraybackslash}p{2.8cm}%
                        >{\raggedright\arraybackslash}p{2.8cm}%
                        >{\raggedright\arraybackslash}p{2.8cm}%
                        >{\raggedright\arraybackslash}p{2.8cm}@{}}
        \toprule
        \textbf{Feature} & \textbf{Proposed Energy Hub} & \textbf{S2-MSNOP \cite{peng}} & \textbf{D-S3NOP \cite{Zhang}} & \textbf{PFCC \cite{purgat2020power}} & \textbf{DPP-Based BESS \cite{DPP}} & \textbf{CFC-DCCB \cite{CFC_DCCB}} \\
        \midrule
        \textbf{Application} 
            & Half-bridge topology for hybrid AC-DC grids 
            & CHB-based medium-voltage multi-feeder systems 
            & Delta-type CHB for medium-voltage multi-feeder (extended range) 
            & Full-bridge topology for low-voltage DC microgrids 
            & Half-bridge topology with BESS in low-voltage DC microgrids 
            & Full-bridge topology for high-voltage DC meshed grids \\
        \addlinespace
        \textbf{Grid Support Functions} 
            & Voltage and reactive power regulation, fault recovery, virtual inertia, virtual resistance 
            & Voltage and reactive power regulation 
            & Voltage and reactive power regulation 
            & Voltage regulation 
            & DC bus regulation, state-of-charge balancing 
            & Fault isolation, line current balancing \\
        \addlinespace
        \textbf{Ripple and Capacitor Management} 
            & Active ripple suppression with film capacitors 
            & Not addressed 
            & Not addressed 
            & Not addressed 
            & Not addressed 
            & Not addressed \\
        \addlinespace
        \textbf{Impedance Measurement and Fault Handling} 
            & Real-time monitoring and bypass, cooperative fault detection and recovery 
            & Not supported 
            & Not supported 
            & Not supported 
            & Not supported 
            & Integrated fault protection, no impedance measurement \\
        \addlinespace
        \textbf{Cost Efficiency} 
            & Low-voltage switches, cost-effective for AC and DC grids 
            & Line-voltage switches 
            & Line-voltage switches 
            & Low-voltage switches 
            & Low-voltage switches 
            & Low-voltage switches \\
        \bottomrule
    \end{tabular}%
    }
\end{table*}

\section{Results}
\subsection{Simulation Results}
\begin{figure*}[htbp]
    \centering
         \begin{minipage}[b]{0.24\textwidth}
        \centering
        \includegraphics[width=\textwidth]{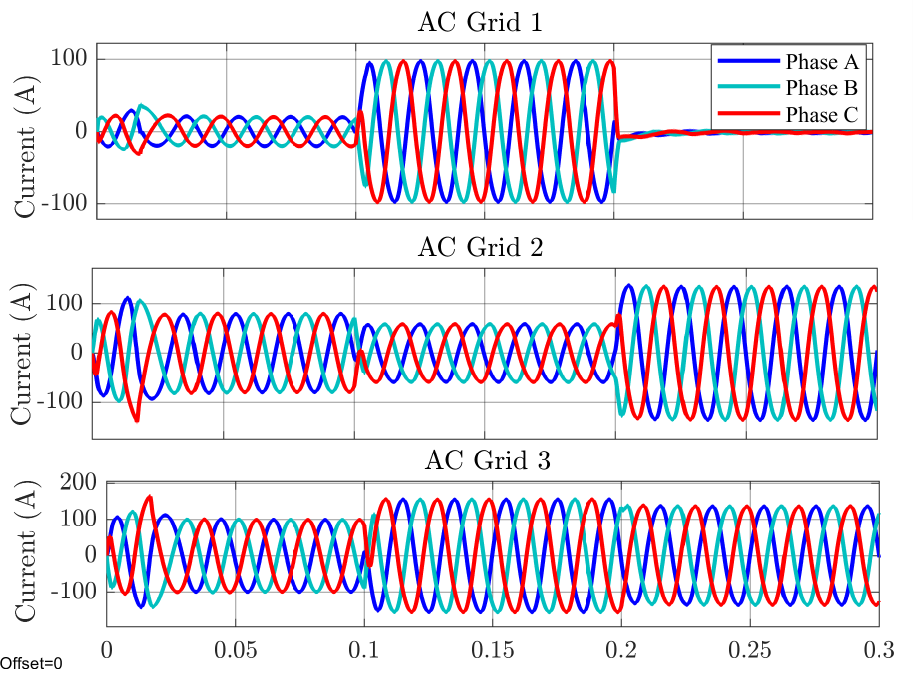}
        \caption{Dynamic power reference tracking in AC grids.}
        \label{fig:Dynamic power reference tracking in AC grids}
    \end{minipage}
    \hfill
    \begin{minipage}[b]{0.24\textwidth}
        \centering
        \includegraphics[width=\textwidth]{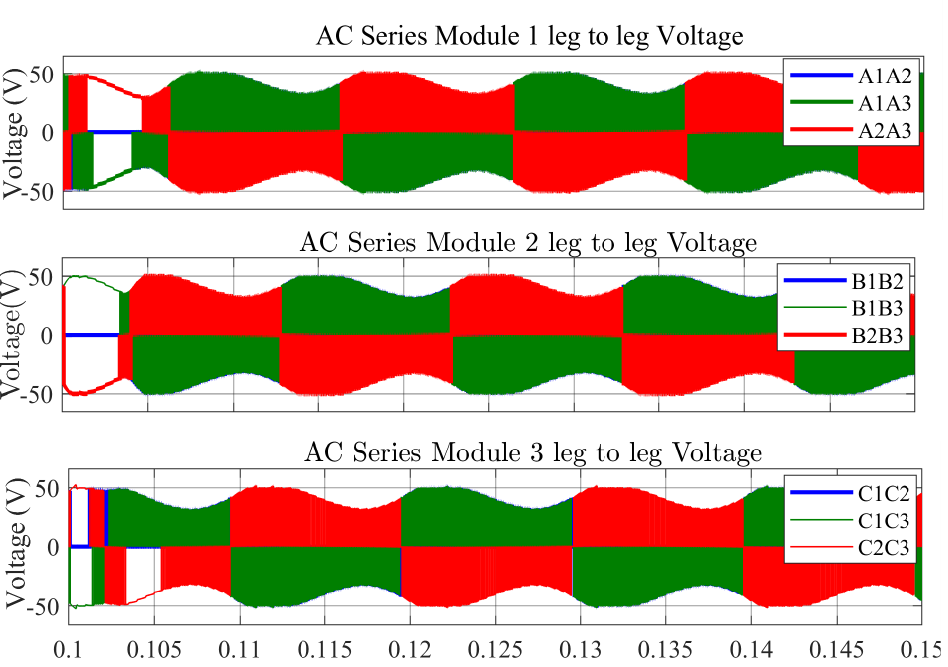}
        \caption{Leg-to-leg voltage of series module, showing low-voltage operation.}
        \label{fig:leg_to_leg_voltage}
    \end{minipage}
    \hfill
    \begin{minipage}[b]{0.24\textwidth}
        \centering
        \includegraphics[width=\textwidth]{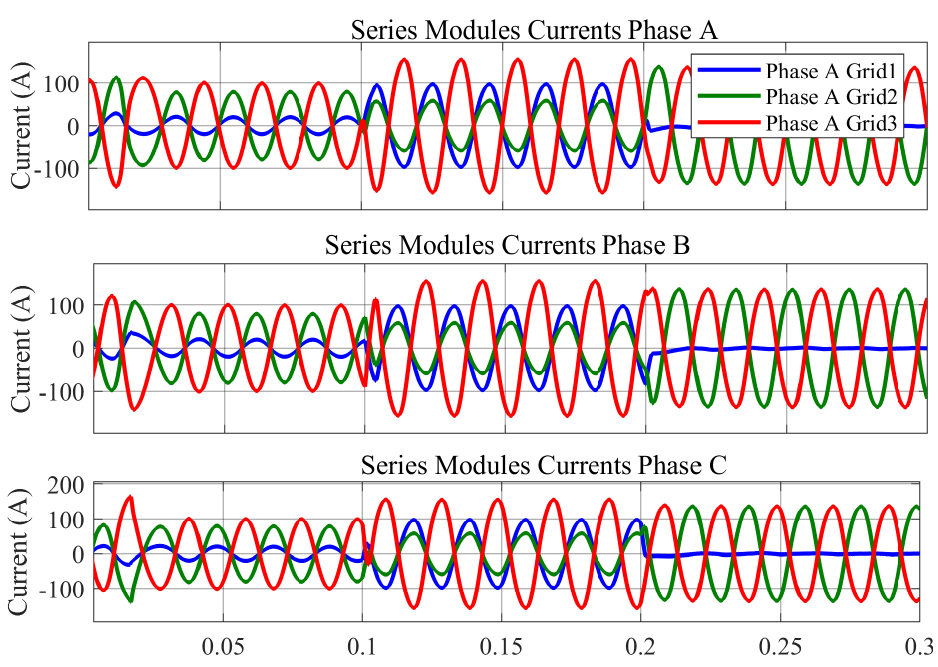}
        \caption{Current waveforms of series modules.}
        \label{fig:series_modules_currents}
    \end{minipage}
    \hfill
      \begin{minipage}[b]{0.24\textwidth}
        \centering
            \includegraphics[width=\textwidth]{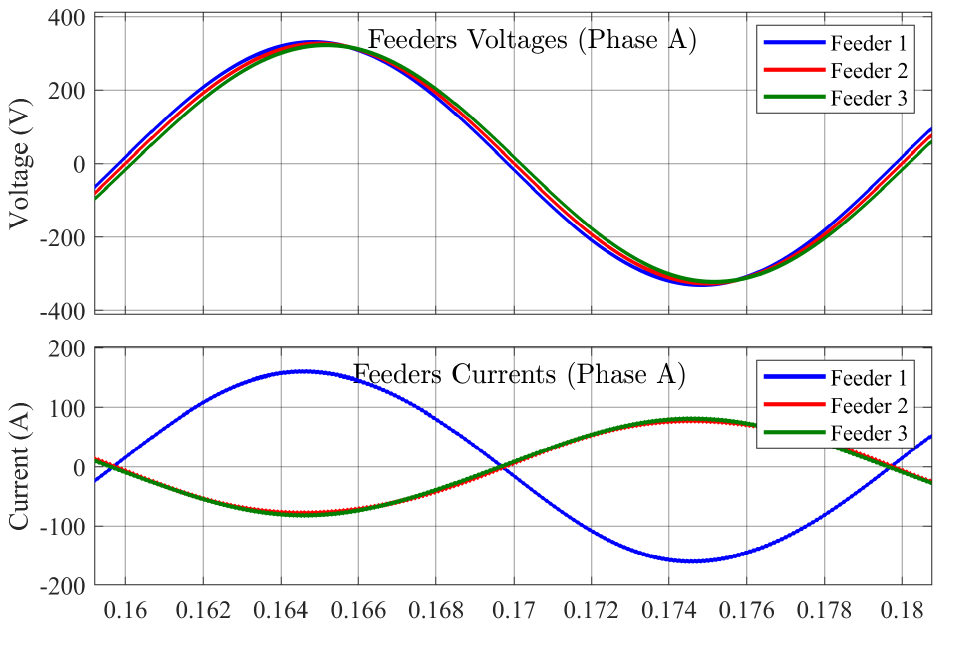}

        \caption{Voltage and Phase Mismatch Mitigation with SMs}
        \label{fig:phase and mag compensation with series module1}
    \end{minipage}
    \hfill
      \begin{minipage}[b]{0.24\textwidth}
        \centering
        \includegraphics[width=\textwidth]{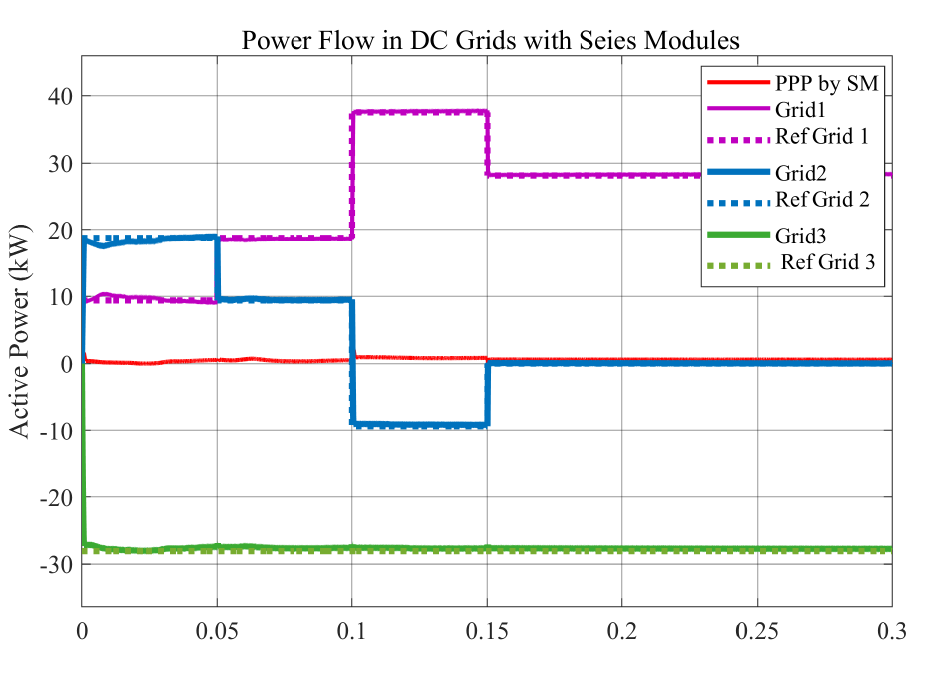}
        \caption{Power flow regulation in DC grid with series modules.}
        \label{fig:dc_powerflow}
    \end{minipage}
      \hfill
    \begin{minipage}[b]{0.24\textwidth}
        \centering
        \includegraphics[width=\textwidth]{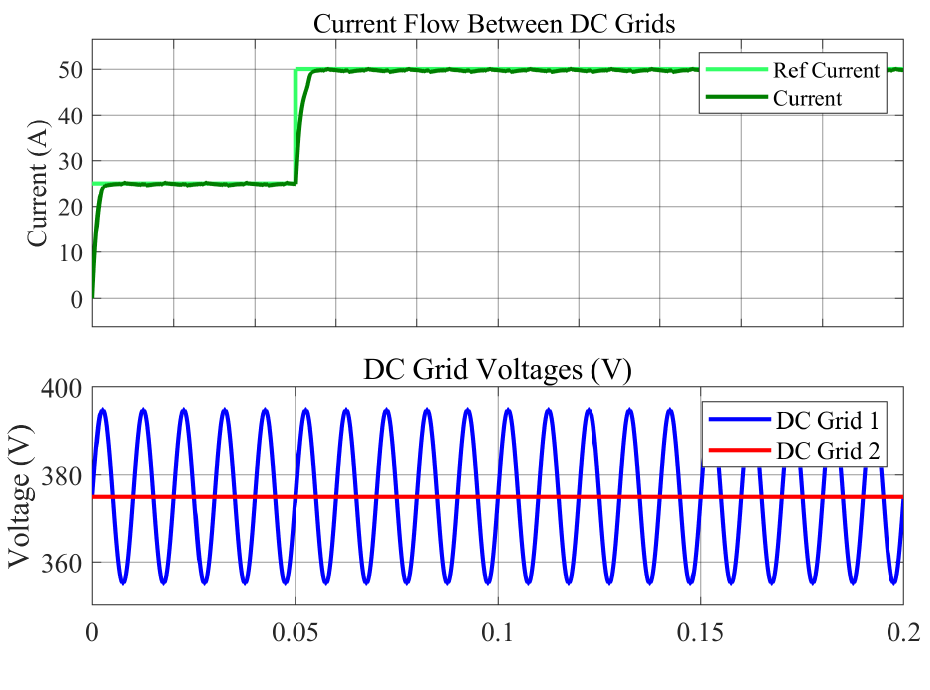}
        \caption{100\,Hz ripple mitigation in DC microgrid with series module.}
        \label{fig:ripple_mitigation}
    \end{minipage}
       \hfill
    \begin{minipage}[b]{0.24\textwidth}
        \centering
        \includegraphics[width=\textwidth]{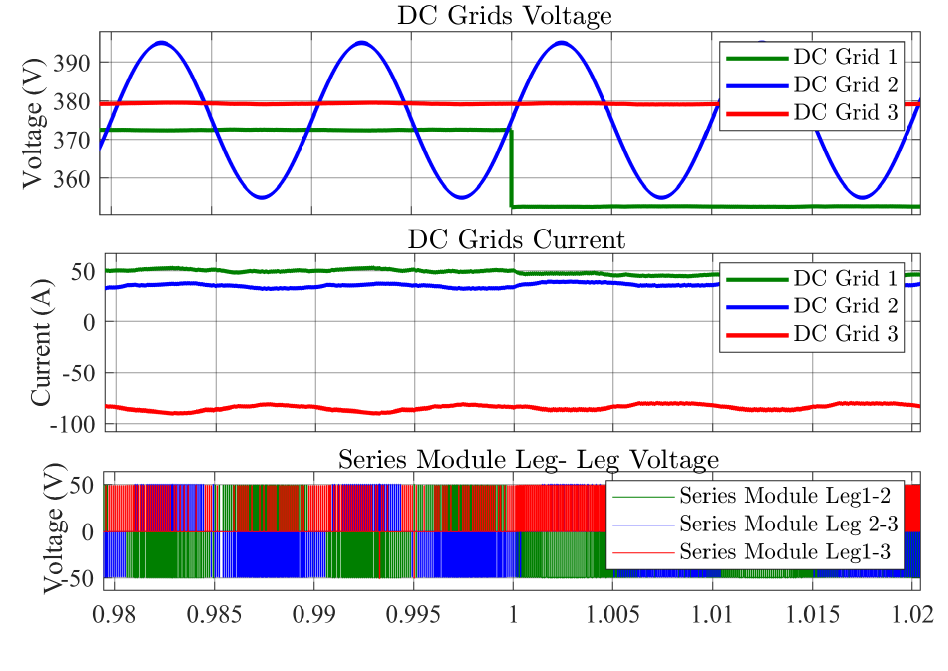}
        \caption{100\,Hz ripple mitigation in DC microgrid with series module.}
        \label{fig:ripple_mitigation2}
    \end{minipage}
     \hfill
       \begin{minipage}[b]{0.24\textwidth}
        \centering
         \includegraphics[width=\textwidth]{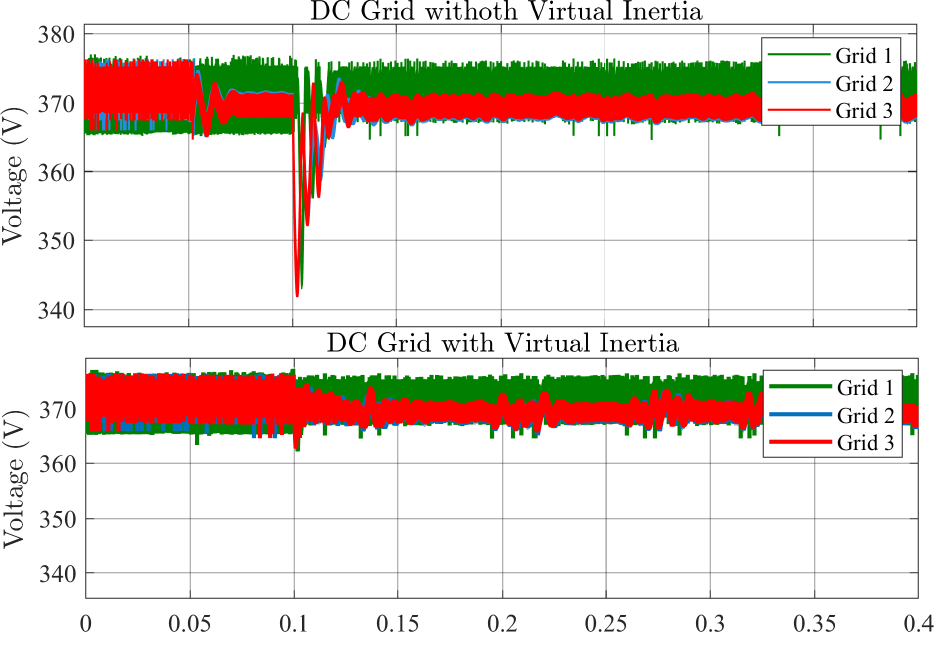}

        \caption{Voltage regulation in DC grids with constant power load, sudden change at $t=0.05$\,s: 
        series modules, and series modules with virtual inertia.}
        \label{fig:virtual_inertia}
    \end{minipage}
          \vspace{0.2cm}
        \begin{minipage}[b]{0.24\textwidth}
        \centering
        \includegraphics[width=\textwidth]{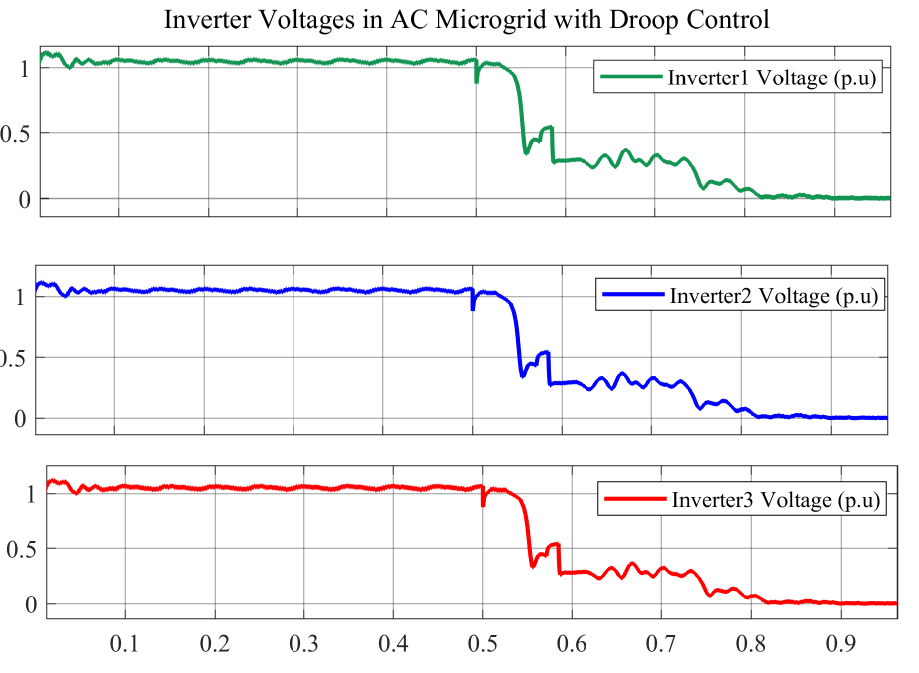}
        \includegraphics[width=\textwidth]{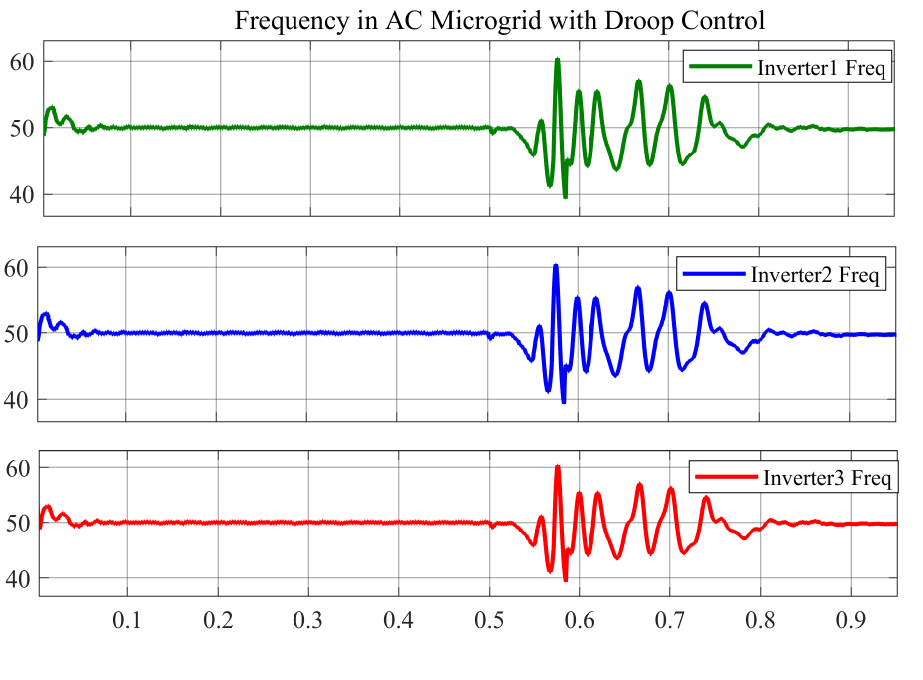}
        \caption{AC microgrid with droop control, three inverter-based grids, 50\% load increase at $t=0.5$\,s.}
        \label{fig:droop_ac}
    \end{minipage}
    \hfill
    \begin{minipage}[b]{0.24\textwidth}
        \centering
        \includegraphics[width=\textwidth]{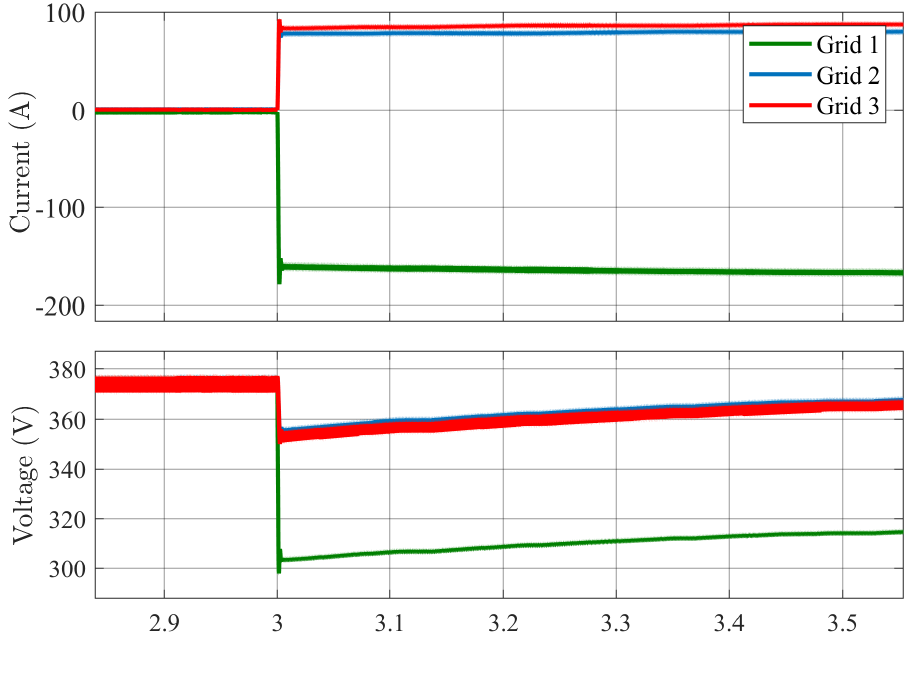}
        \includegraphics[width=\textwidth]{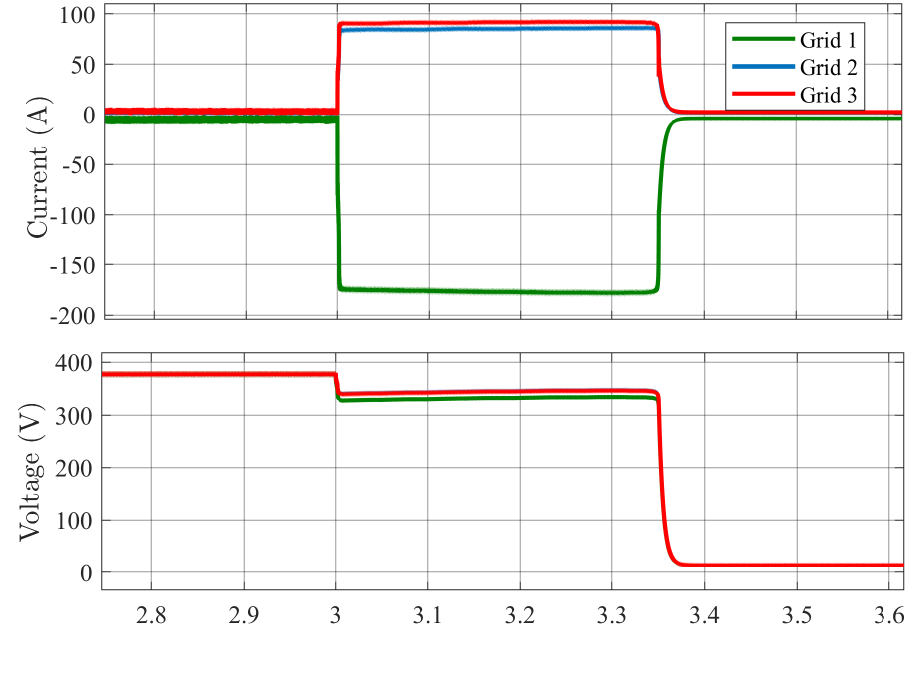}
        \caption{Current sharing and voltage regulation in DC grid with series modules, droop control, load step at $t=3$\,s.}
        \label{fig:droop_SM_DC}
    \end{minipage}
    \hfill
    \begin{minipage}[b]{0.24\textwidth}
        \centering
        \includegraphics[width=\textwidth]{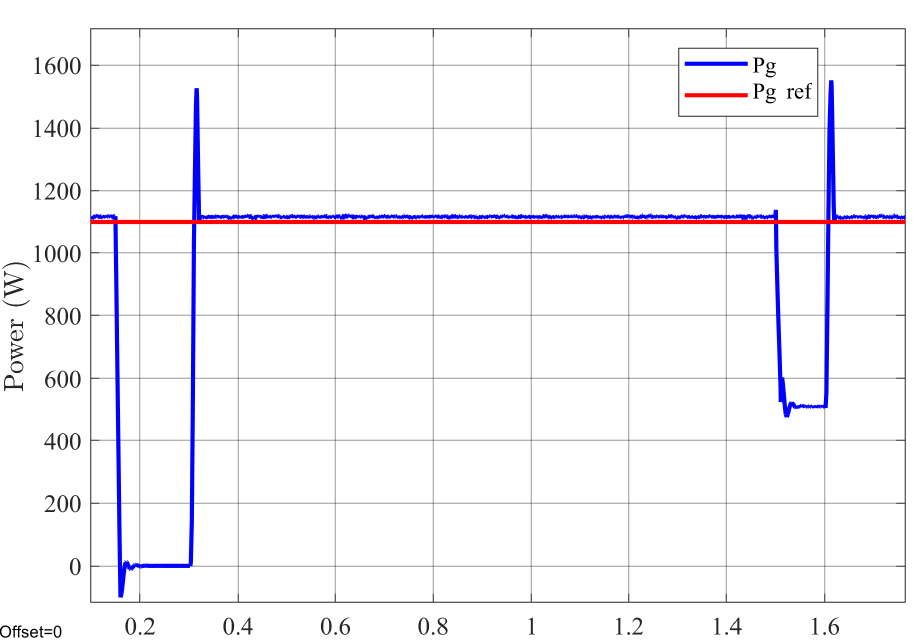}
        \includegraphics[width=\textwidth]{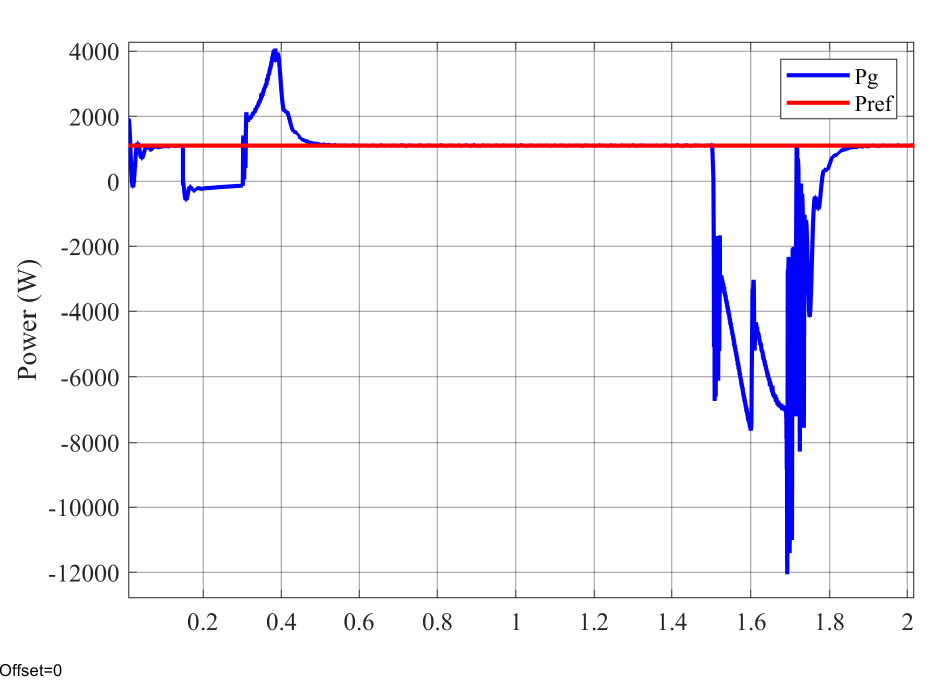}
        \caption{Grid-forming inverter power injection to weak grid, with series modules (top) and conventional setup (bottom), under three-phase short circuit (0.15--0.3\,s) and impedance shift (1.5--1.6\,s).}
        \label{fig:GF_conventional}
    \end{minipage}
     \hfill
      \begin{minipage}[b]{0.24\textwidth}
        \centering
        \includegraphics[width=\textwidth]{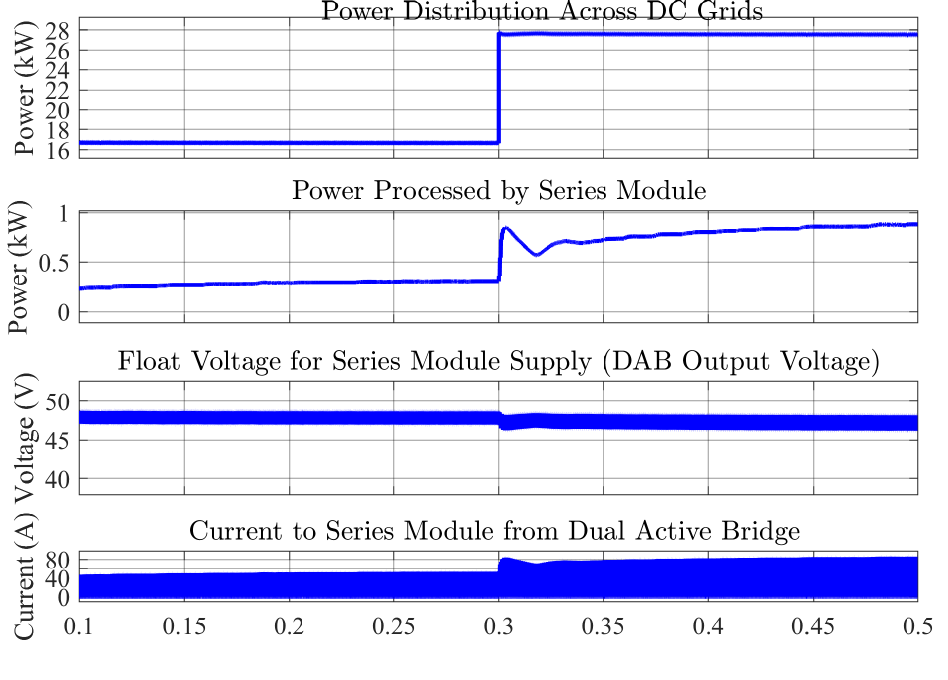}
        \includegraphics[width=\textwidth]{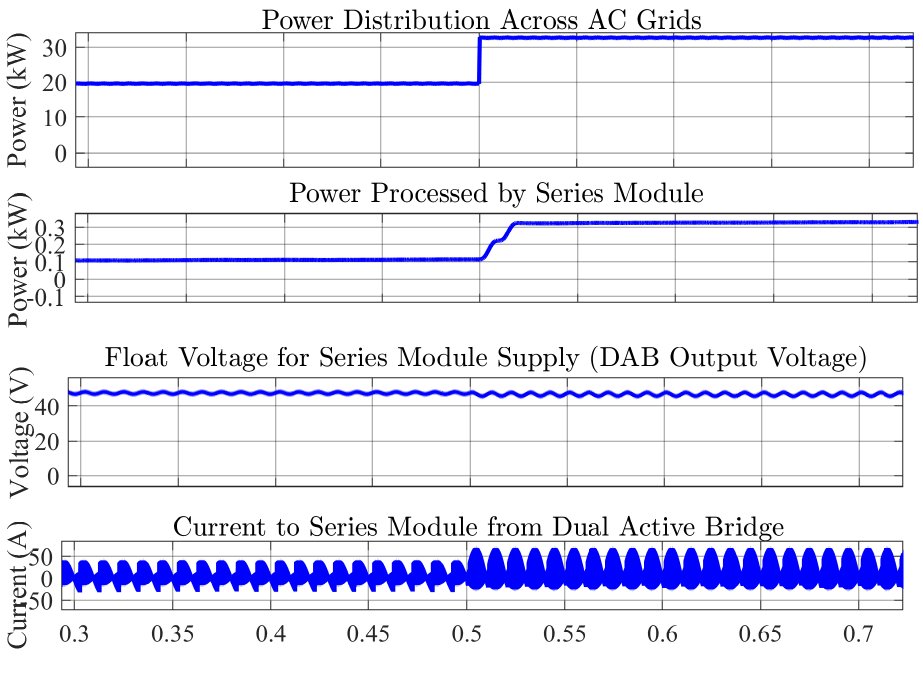}
        \caption{Grid Power Control via Partial Power Processing (less Power to control the whole power between Grids)(Grid Power Control with Minimal Processing)}
        \label{fig:PPP}
    \end{minipage}
   \caption{Simulation results for power flow, voltage regulation, and current sharing in AC and DC grids.}
    \label{fig:simulation_results}
\end{figure*}

We evaluated the energy router architecture using MATLAB/Simulink and OPAL-RT simulations to assess its dynamic performance and control strategies across diverse operating conditions for AC and DC grids. Unlike conventional droop control, which exhibits instability during sudden load changes and voltage/impedance variations (Figs.~\ref{fig:droop_ac}, \ref{fig:droop_SM_DC}, 14, and 15), the proposed system ensures robust stability and performance. Key findings include:

\begin{itemize}
    \item \textbf{Power Flow Regulation}: The router maintained stable power flow between AC and DC grids, effectively responding to load changes and impedance variations, as shown in Figs.~7, 9, 11, 21, 22, and 23.
    \item \textbf{Voltage and Phase Compensation}: Series modules corrected voltage magnitude and phase mismatches, ensuring stable power flow across grids (Fig.~10).
    \item \textbf{Ripple Mitigation}: The proposed system actively suppresses DC grid voltage ripple without relying on large passive components, as demonstrated in Figs. 10, 11, and 22. In Figs. 10 and 22, the system maintains stability while adjusting the transmitted power reference between two DC feeders, where one feeder exhibits a 100 Hz ripple with a 5\% amplitude. In Fig. 11, the system stabilizes a three-terminal DC network, where one feeder experiences a 100 Hz ripple with a 5\% amplitude and a separate feeder sustains a 5\% voltage drop.
    \item \textbf{Voltage Regulation under Constant Power Loads}: 
Figure 12 illustrates the voltage regulation performance in a DC grid with a constant power load, subjected to a sudden change at $ t = 0.05 \, \text{s} $. The response of series modules alone and series modules with virtual inertia is compared, highlighting their effectiveness in maintaining stable voltage levels.
        
    \item \textbf{Grid-Forming Inverter Performance}: The grid-forming inverter, paired with series modules, showed enhanced fault recovery and stability during short circuits and impedance shifts compared to conventional systems (Fig.~15, 25).
\end{itemize}

\begin{figure*}[htbp]
    \centering
    \begin{minipage}[b]{0.32\textwidth}
        \centering
        \includegraphics[width=\textwidth]{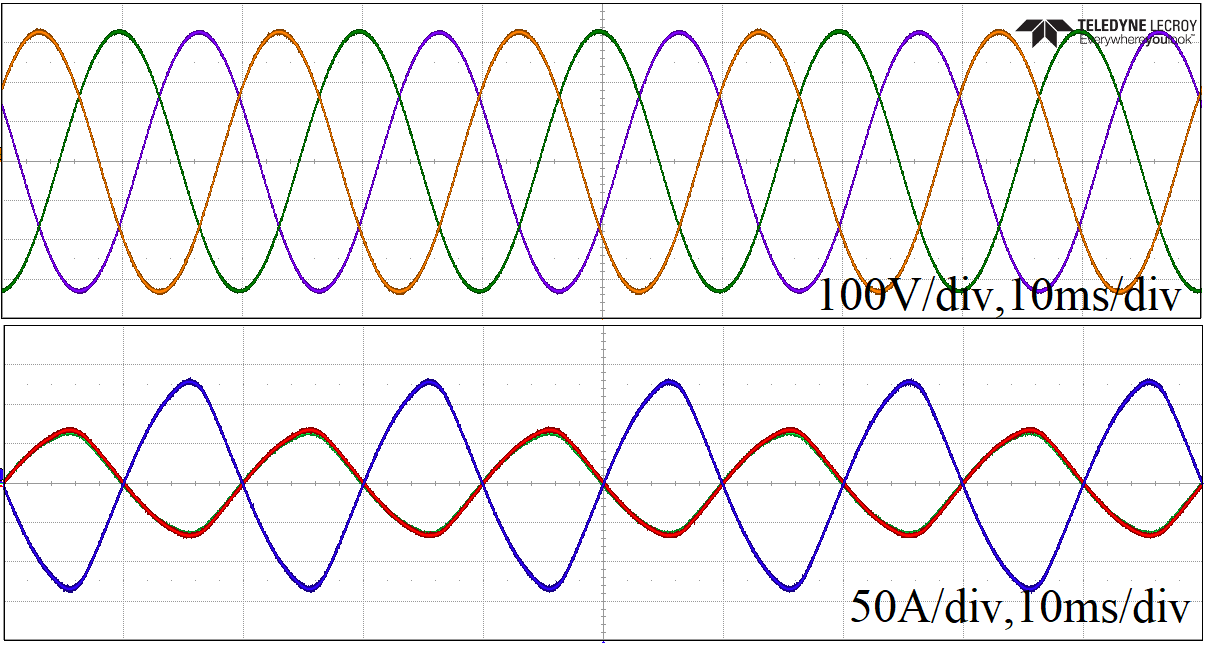}
        \caption{AC power flow initial state: Grid Voltages (top) and Series Module currents (bottom) }
        \label{fig:ac_power_flow_opal}
    \end{minipage}
    \hfill    
        \begin{minipage}[b]{0.32\textwidth}
        \centering            \label{fig:ac_power_flow_transition}
    \end{minipage}
    \hfill
    \begin{minipage}[b]{0.33\textwidth}
        \centering
        \label{fig:ac_power_flow_Vph-ph_SM_initital}
    \end{minipage}
    \hfill
        \begin{minipage}[b]{0.32\textwidth}
        \centering
       \includegraphics[width=\textwidth]{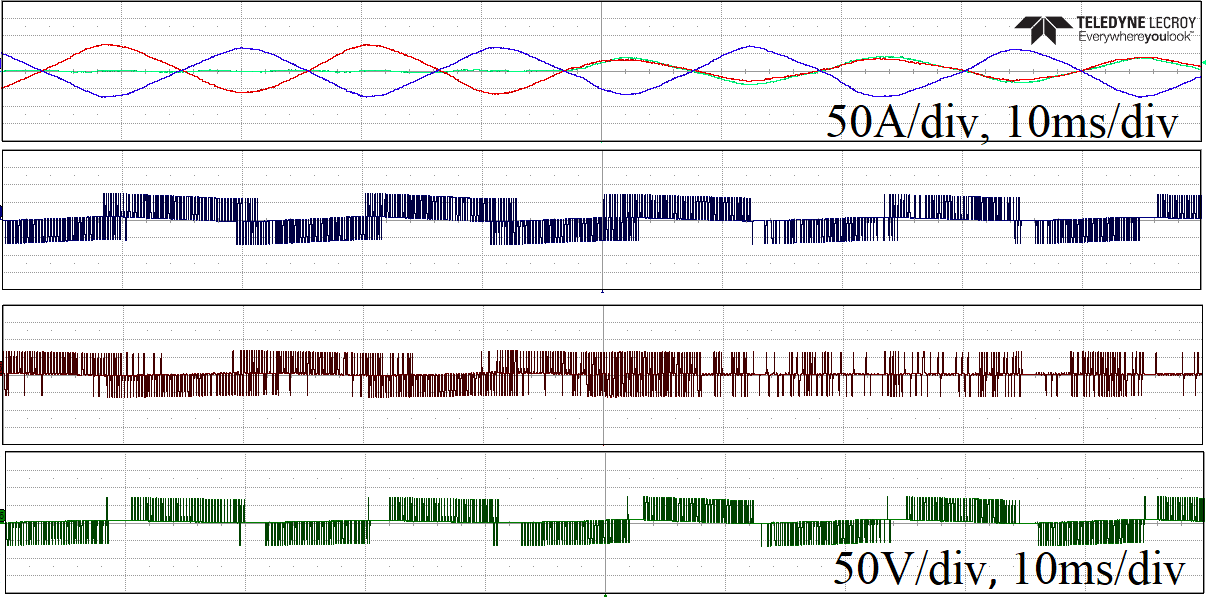}      
        \caption{AC power flow transient state: Series module currents (top) and voltages between series module A legs (bottom).}
        \label{fig:ac_power_flow_Vph-ph_SM_transition}
    \end{minipage}
    \hfill
    \begin{minipage}[b]{0.33\textwidth}
        \centering        \label{fig:ac_power_flow_Vfeeders_grids}
    \end{minipage}
    \hfill
    \begin{minipage}[b]{0.32\textwidth}
        \centering
        \label{fig:ac_transition}
    \end{minipage}
    \hfill
    \begin{minipage}[b]{0.32\textwidth}
        \centering
        \includegraphics[width=\textwidth]{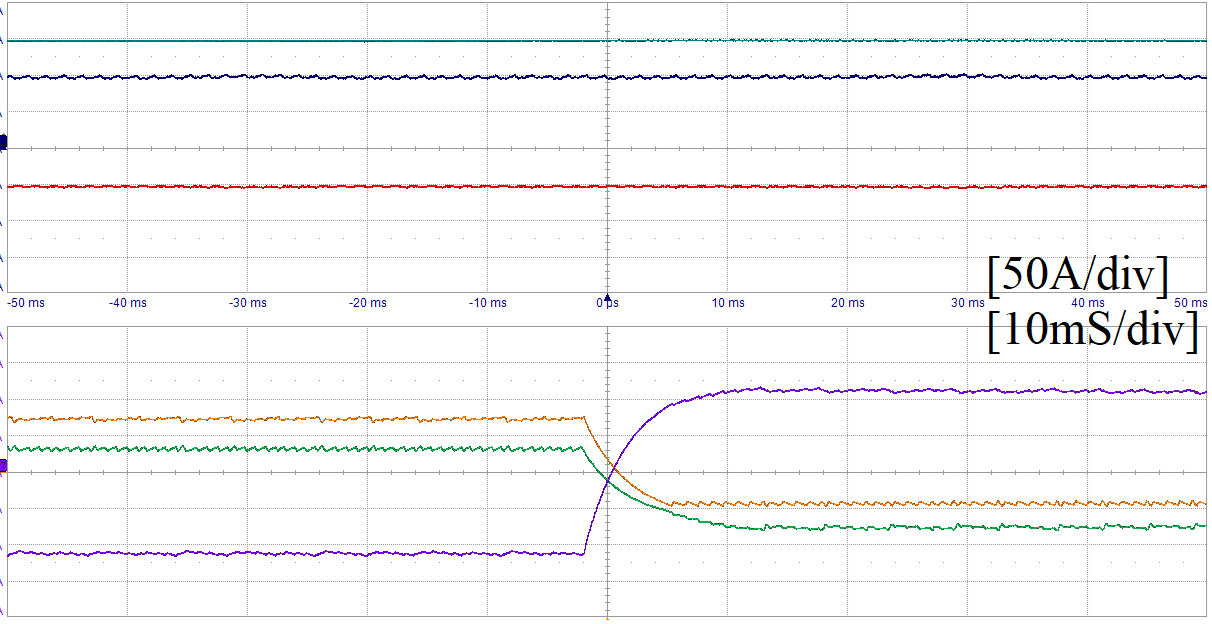}
        \caption{Bipolar DC Grid Current Flow: Positive Pole (Top), Negative Pole (Bottom).}
        \label{fig:dc_transition1}
    \end{minipage}
    \vspace{0.2cm}
\begin{minipage}[b]{0.32\textwidth}
        \centering
        \includegraphics[width=\textwidth]{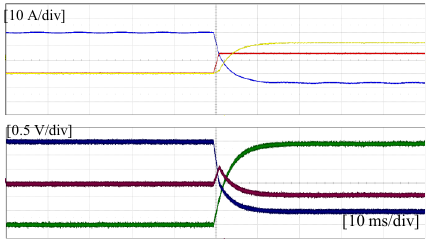}
        \caption{ DC Grid Current Flow: Positive Pole (Top), DC Series Module leg-leg Voltages.}
        \label{fig:dc_transition_droop}
    \end{minipage}
    \hfill
    \begin{minipage}[b]{0.33\textwidth}
        \centering
        \includegraphics[width=\columnwidth]{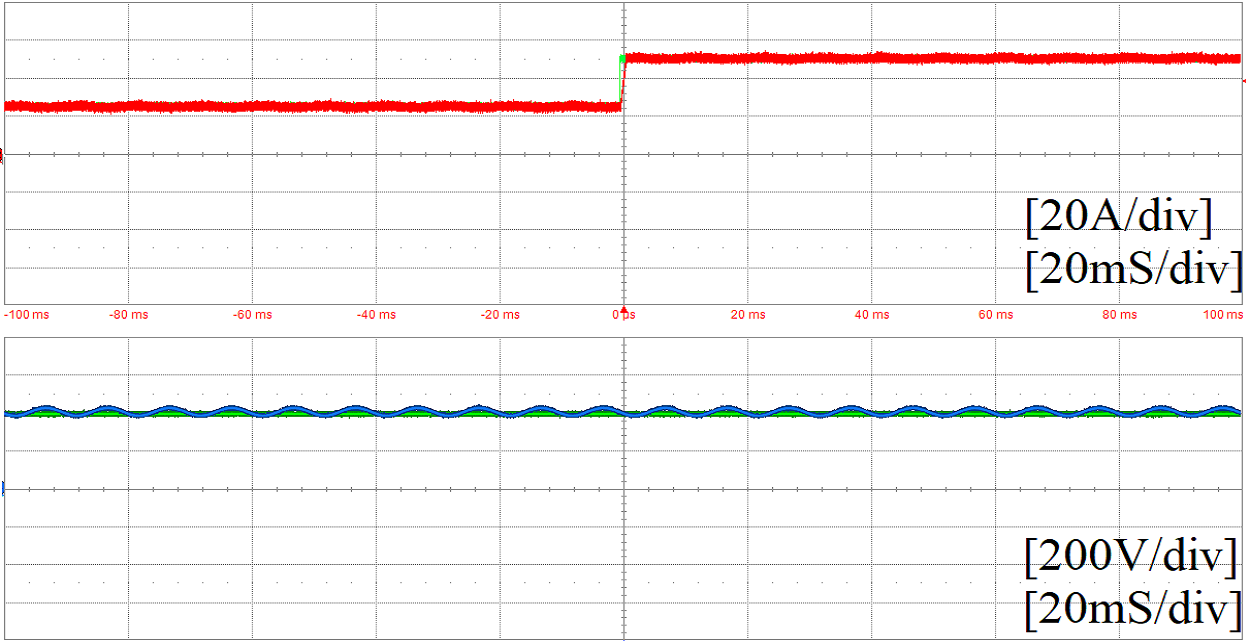}
        \caption{DC grid with 100\,Hz ripple, series modules suppress ripple transfer.}
        \label{fig:ripple_mitigation_opal}
    \end{minipage}
        \hfill
    \begin{minipage}[b]{0.32\textwidth}
        \centering
        \includegraphics[width=\textwidth]{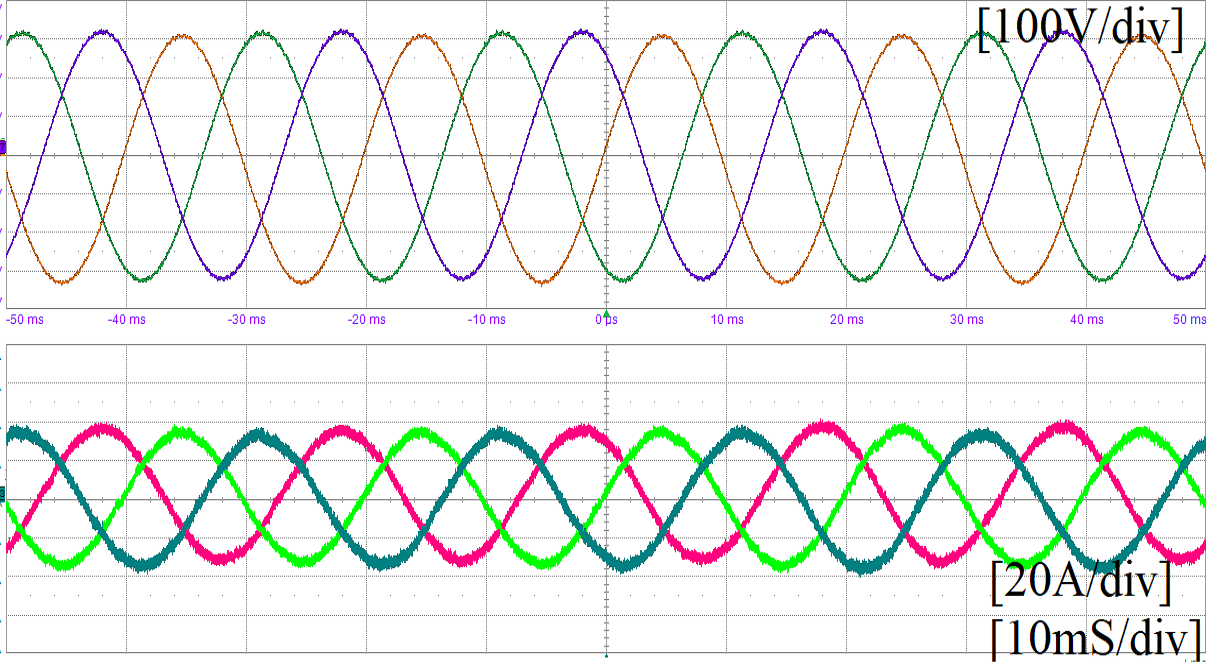}
        \caption{AFE waveforms: voltage (top) and current (bottom).}
        \label{fig:ac_stable}
    \end{minipage}

              \hfill
    \begin{minipage}[b]{0.32\textwidth}
        \centering
        \includegraphics[width=\columnwidth]{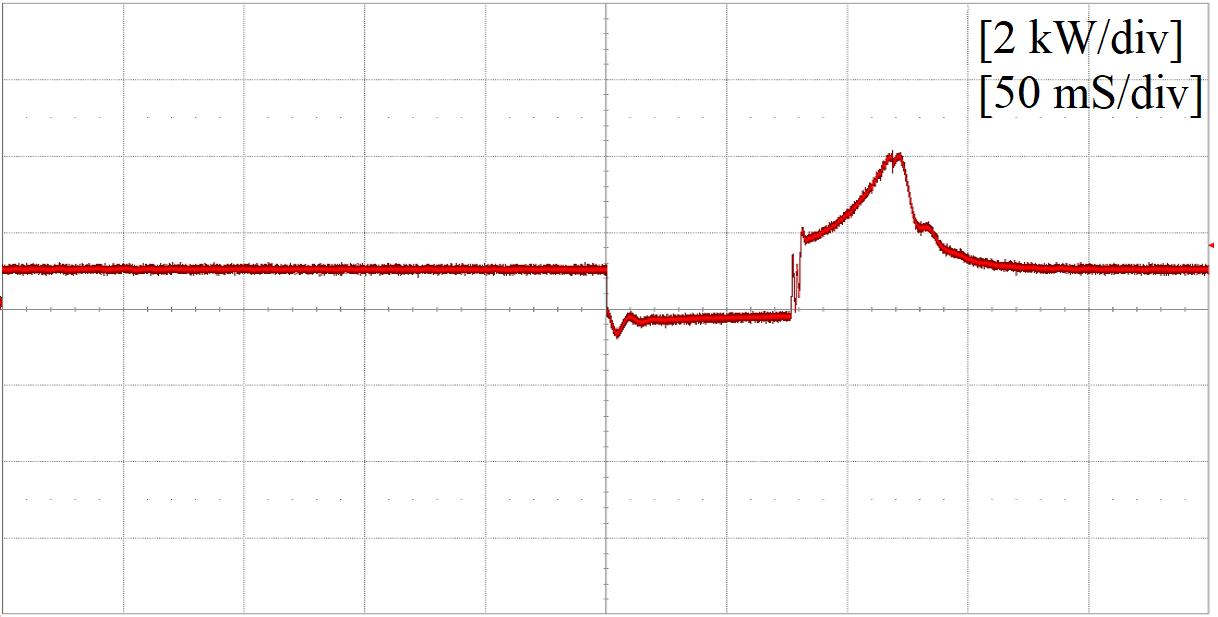}
        \caption{Power injected by the grid-forming inverter in a conventional setup during a three-phase short-circuit (t = 1.00–1.15 s)}
        \label{fig:GF_Conventional_opal}
    \end{minipage}
    \hfill
    \begin{minipage}[b]{0.32\textwidth}
        \centering
        \includegraphics[width=\columnwidth]{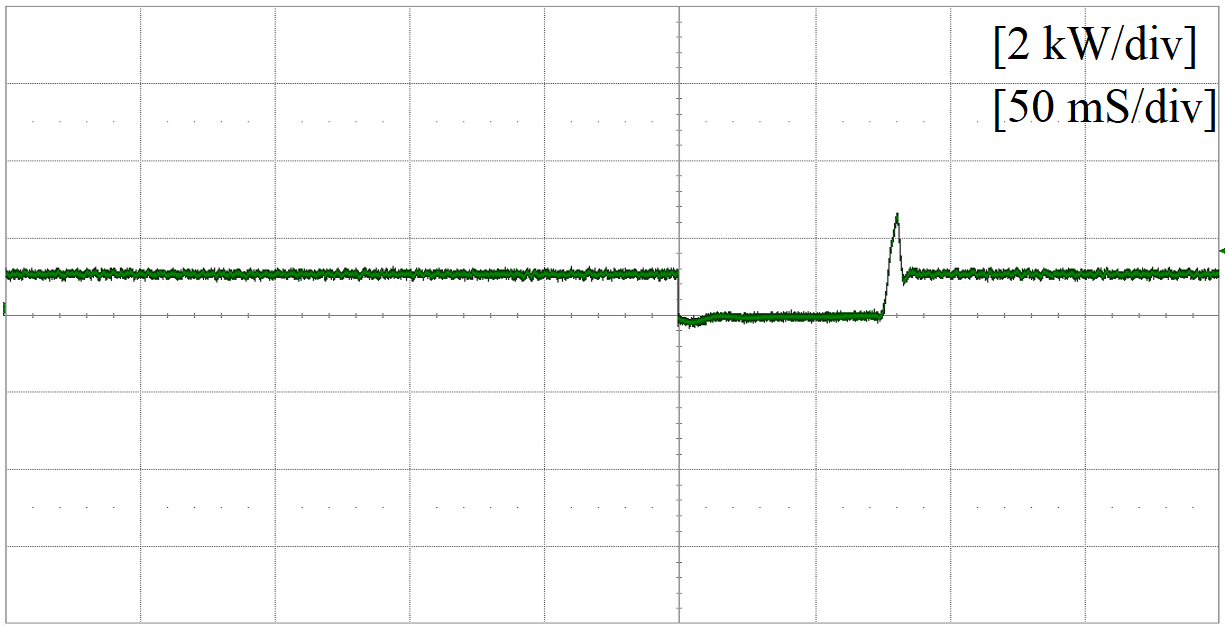}
        \caption{Power injected into the grid by the grid-forming inverter with series modules during a three-phase short-circuit (t = 1.00–1.15 s).}
        \label{fig:GF_SM_opal}
    \end{minipage}
    \hfill
\begin{minipage}[b]{0.32\textwidth}
\centering
\includegraphics[width=\textwidth]{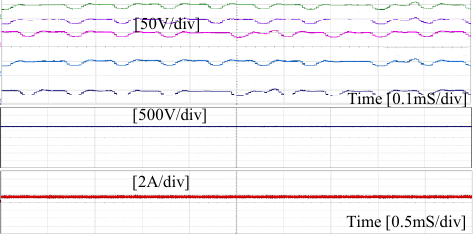}
\caption{DAB voltages (top) and AFE DC-side voltage/current (bottom).}
\label{fig:DABsAFE}
\end{minipage}

    \caption{Real-time Opal simulation results validating control strategy in AC/DC systems.}    
\label{fig:OPAL_RT Simulations}
\end{figure*}

\subsection{Experimental Setup and Results}

\begin{figure}[t]
    \centering
    \includegraphics[width=0.8\columnwidth]{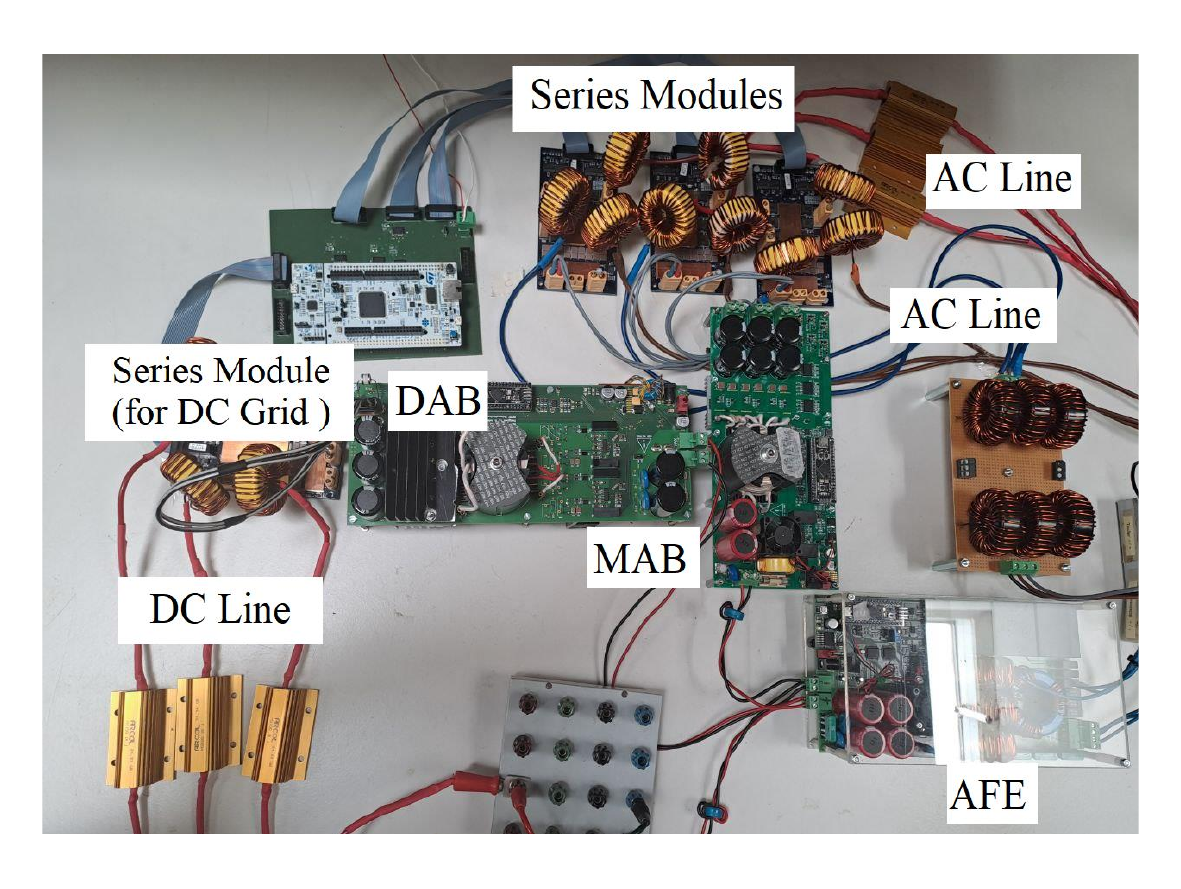}
    \caption{Experimental prototype}
    \label{fig:prototype}
\end{figure}

\begin{figure}[t]
    \centering
    \includegraphics[width=0.8\columnwidth]{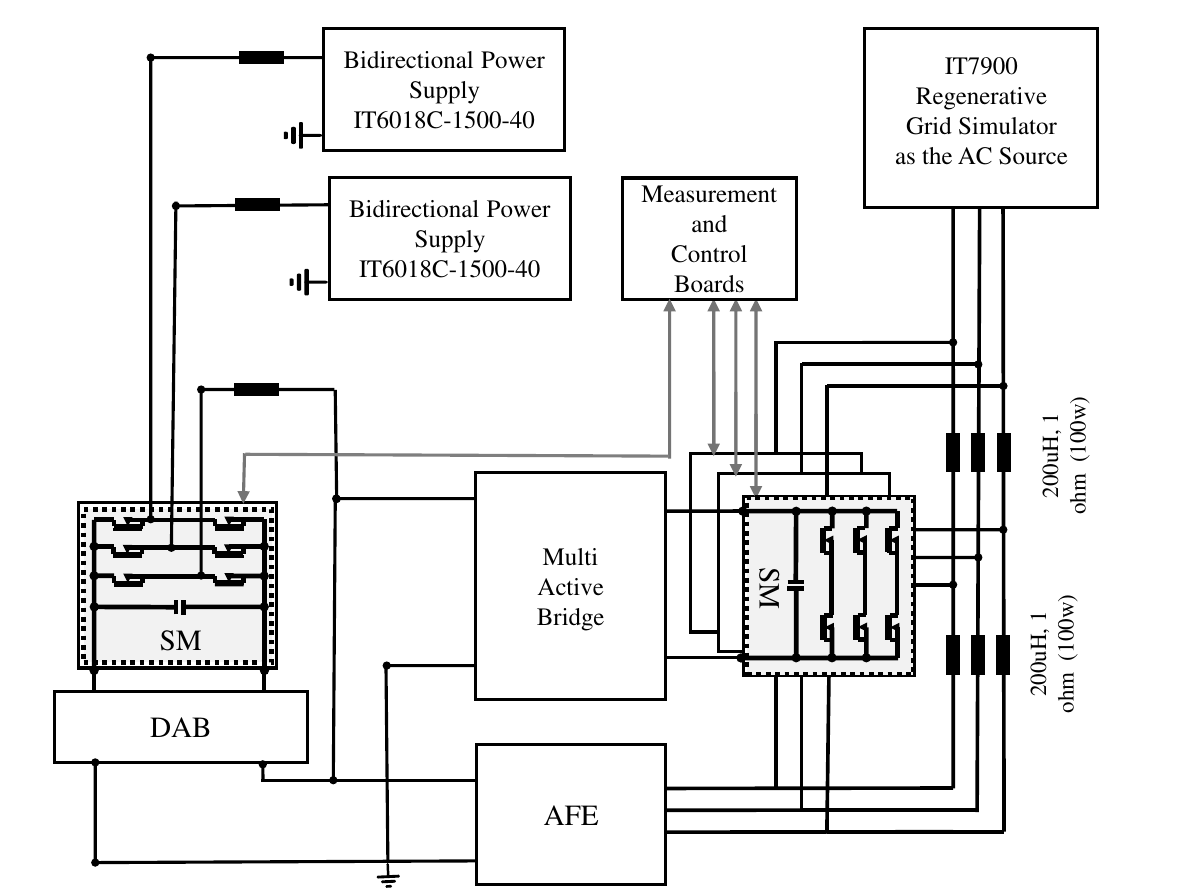}
    \caption{Structure of the experimental test bench}
    \label{fig:schematic}
\end{figure}

\begin{figure*}[htbp]


\hfill

\begin{minipage}[b]{0.32\textwidth}
\centering
\includegraphics[width=\textwidth]{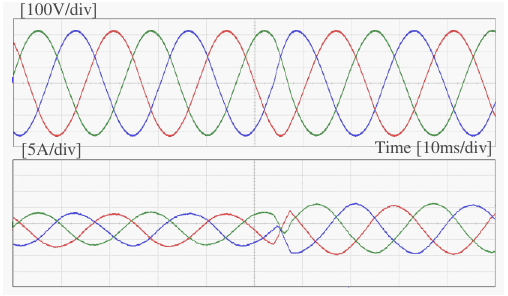}
\caption{Feeder~A voltages (top) and currents (bottom) during power reversal.}
\label{fig:GridTransient}
\end{minipage}
\hfill
\begin{minipage}[b]{0.32\textwidth}
\centering
\includegraphics[width=\textwidth]{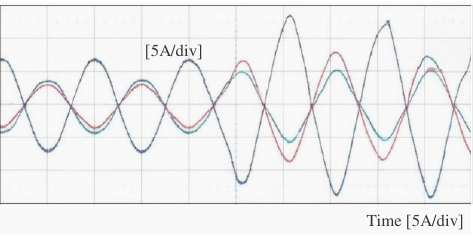}
\caption{AC series-module leg currents during the transition.}
\label{fig:ACtransition}
\end{minipage}
\hfill
\begin{minipage}[b]{0.32\textwidth}
\centering
\includegraphics[width=\textwidth]{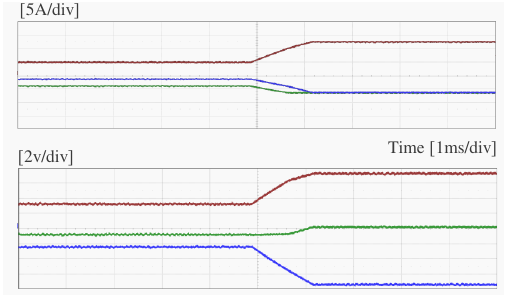}
\caption{DC grid current flow (top) and DC series-module leg-to-leg voltages (bottom).}
\label{fig:DC1}
\end{minipage}

\hfill
\begin{minipage}[b]{0.19\textwidth}
\centering
\includegraphics[width=\textwidth]{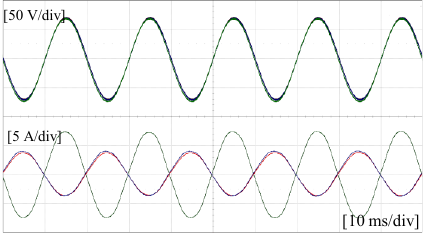}
\caption{$P$ injected into Feeder~3 (others slack).}
\label{fig:Sc1}
\end{minipage}
\hfill
\begin{minipage}[b]{0.19\textwidth}
\centering
\includegraphics[width=\textwidth]{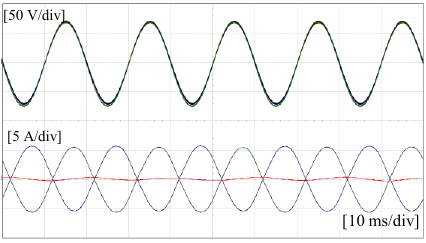}
\caption{$P$ into Feeder~3, Feeder~2 power kept at zero.}
\label{fig:Sc2}
\end{minipage}
\hfill
\begin{minipage}[b]{0.19\textwidth}
\centering
\includegraphics[width=\textwidth]{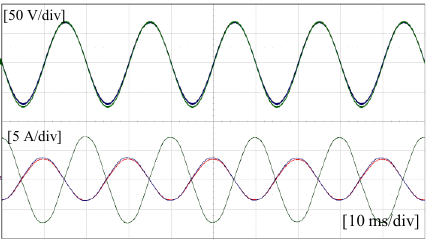}
\caption{$Q$ injected into Feeder~3 (others slack).}
\label{fig:Sc3}
\end{minipage}
\hfill
\begin{minipage}[b]{0.19\textwidth}
\centering
\includegraphics[width=\textwidth]{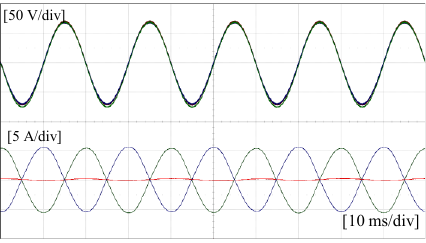}
\caption{$Q$ into Feeder~3, Feeder~2 power kept at zero.}
\label{fig:Sc4}
\end{minipage}
\hfill
\begin{minipage}[b]{0.19\textwidth}
\centering
\includegraphics[width=\textwidth]{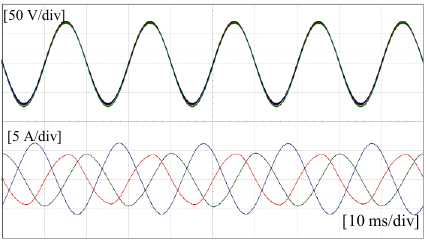}
\caption{ simultaneous $P$ into Feeder~2 and $Q$ into Feeder~3.}
\label{fig:Sc5}
\end{minipage}

\hfill
\begin{minipage}[b]{0.19\textwidth}
\centering
\includegraphics[width=\textwidth]{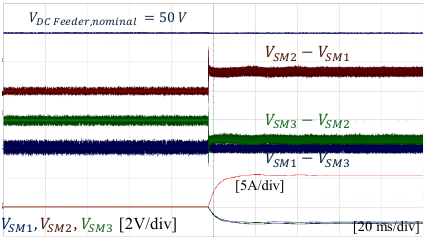}
\caption{ $P$ injected into Feeder~2, Feeder~1 and Feeder~3 are slack.}
\label{fig:Sc1dc}
\end{minipage}
\hfill
\begin{minipage}[b]{0.19\textwidth}
\centering
\includegraphics[width=\textwidth]{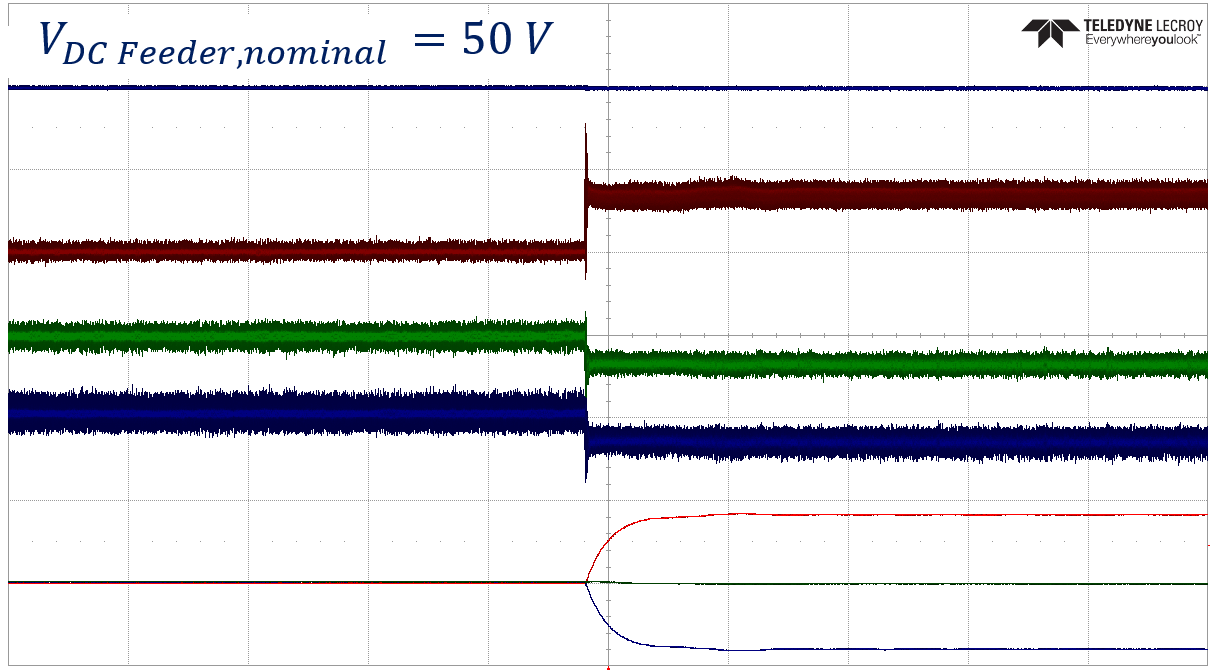}
\caption{$P$ into Feeder~2, Feeder~3 power kept at zero,  and the remaining feeder~1 is slack.}
\label{fig:Sc2dc}
\end{minipage}
\hfill
\begin{minipage}[b]{0.19\textwidth}
\centering
\includegraphics[width=\textwidth]{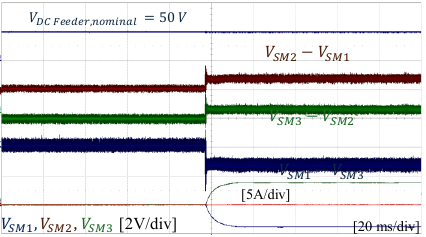}
\caption{ $P$ into Feeder~3, Feeder~2 power kept at zero,  and Feeder~1 is slack. }
\label{fig:Sc4dc}
\end{minipage}
\hfill
\begin{minipage}[b]{0.19\textwidth}
\centering
\includegraphics[width=\textwidth]{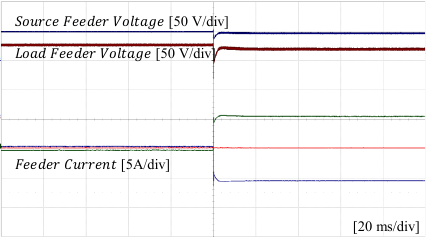}
\caption{Load connected to Feeder~3 without the series module: the feeder exhibits load-induced voltage drop.}
\label{fig:Sc6dc}
\end{minipage}
\hfill
\begin{minipage}[b]{0.19\textwidth}
\centering
\includegraphics[width=\textwidth]{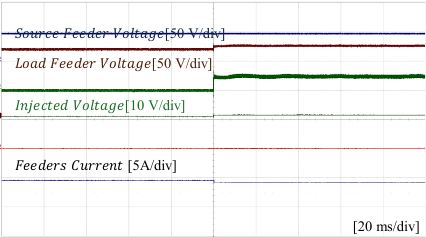}
\caption{Voltage stabilization on Feeder~3 with the series module despite the load changing.}
\label{fig:Sc7dc}
\end{minipage}

\caption{Experimental results for AC/DC Energy Router operation: voltage/phase alignment, dynamic transients, selective decoupled $P/Q$ control across feeders, and Voltage stabilization with Series Modules.}
\label{fig:experimental_results}
\end{figure*}

The Energy Router was validated on a hardware test bench that emulates a multi-port hybrid AC/DC hub (Figs.~\ref{fig:prototype}, \ref{fig:schematic}). An IT7900 regenerative source supplied three three-phase AC feeders through line impedances of 200~µH + 1~$\Omega$. Each feeder interfaced with a dedicated series module including an additional 250~µH inductor. On the DC side, two IT6018C-1500-40 bidirectional supplies represented independent DC grids. An active front end (AFE) generated a 400~V DC hub and powered the isolated floating supplies (50~V) of the series modules via DABs (DC modules) and an MAB (AC modules). Each module used IPT015N10N5 MOSFETs ($r_{\mathrm{DS(on)}}<2$~m$\Omega$) and a 300~µF film capacitor array. LV~25-P and LA~55-P sensors (10~kHz) provided measurements to an ARM controller that executed the proposed control algorithms.
The experimental results demonstrate:  
(i) \textbf{Voltage/phase alignment} between AC feeders;  
(ii) \textbf{Dynamic behavior} during power direction changes, including feeder voltages/currents and series-module leg currents (Figs.~\ref{fig:GridTransient}–\ref{fig:DC1});  
(iii) \textbf{Selective, decoupled power control} across feeders via five scenarios (Figs.~\ref{fig:Sc1}–\ref{fig:Sc5}); 
and  
(iii) \textbf{Selective, decoupled power control} across DC feeders via three power-flow scenarios (Figs.~\ref{fig:Sc1dc}–\ref{fig:Sc4dc}); additionally, two figures verify that injecting power from an idle feeder stabilizes the load-bus voltage (Figs.~\ref{fig:Sc6dc}–\ref{fig:Sc7dc}).
The voltage/phase results verify precise correction of magnitude mismatch and phase error between feeders. The dynamic tests show smooth transients and well-damped responses during active power reversal without jeopardizing synchronization. The five multi-feeder scenarios confirm that the Energy Router tracks per-feeder setpoints while keeping active and reactive power decoupled.


\vspace{2mm}

\section{Conclusion}
This paper introduces a novel energy hub architecture that seamlessly integrates renewable energy sources and high-power devices within hybrid AC and DC grids. By employing modular series modules and a Battery Energy Storage System, the proposed hub achieves precise power flow control, enhanced grid stability, and superior power quality.
The advanced control strategy combines inertia emulation and dynamic power flow regulation to directly counteract rapid frequency deviations while suppressing high-frequency disturbances. The integrated approach ensures effective energy buffering for transient support, leading to improved system damping and robust stability. Additionally, the dynamic regulation of the series modules minimizes ripple voltage and stabilizes DC links, thereby reducing the reliance on bulky passive components and enabling the use of compact, high-reliability film capacitors.
Extensive experimental validations and simulations confirm that the energy hub delivers higher stability margins, enhanced damping, and superior dynamic performance. This innovative integration of control mechanisms and active ripple mitigation not only addresses the challenges of modern grids with high renewable penetration but also lays a strong foundation for future advancements in energy management systems.

\newpage
\vfill

\begin{thebibliography}{99}
\bibliographystyle{IEEEtran}

\bibitem{SGoetz}
M. Lu and S. M. Goetz, ``A Highly Compact Transformerless Universal Power-Flow and Quality Control as well as Soft Open Point Circuit,'' \textit{Annual Conference of the IEEE Industrial Electronics Society, IECON}, Brussels, Belgium, Dec. 2022, vo. 48, doi: 10.1109/IECON49645.2022.9968691.

\bibitem{richardson2010electric}
D. B. Richardson, ``Electric vehicles and the electric grid: A review of modeling approaches, impacts, and renewable energy integration,'' \textit{Renew. Sustain. Energy Rev.}, vol. 19, pp. 247--254, 2013.

\bibitem{sepehrzad2023}
R. Sepehrzad et al., ``Improved Power Sharing and Energy Management Platform in Microgrid Considering Stochastic Dynamic Behavior of the Electric Vehicles,'' \textit{Sustainable Cities and Society}, p. 104826, 2023.

\bibitem{benhassi2024}
Y. A. Ait Ben Hassi, Y. Hennane, and A. Berdai, ``Innovative power sharing and secondary controls for meshed microgrids,'' \textit{Int. J. Elect. Comput. Eng.}, vol. 15, no. 1, pp. 99--113, 2024.

\bibitem{abouassi2023}
A. Abouassi et al., ``Power Sharing Issues and Control Strategies in Islanded Microgrid: A Comparative Study,'' 2023.

\bibitem{mishra2024}
B. Mishra and M. Pattnaik, ``A modified droop-based decentralized control strategy for accurate power sharing in a PV-based islanded AC microgrid,'' \textit{ISA Trans.}, vol. 153, pp. 467--481, 2024.

\bibitem{payghami}
S. Peyghami, P. Davari, H. Mokhtari and F. Blaabjerg, "Decentralized Droop Control in DC Microgrids Based on a Frequency Injection Approach," in IEEE Transactions on Smart Grid, vol. 10, no. 6, pp. 6782-6791, Nov. 2019, doi: 10.1109/TSG.2019.2911213.

\bibitem{mokhtari}
M. Jafari, S. Peyghami, H. Mokhtari and F. Blaabjerg, "Enhanced Frequency Droop Method for Decentralized Power Sharing Control in dc Microgrids," in IEEE Journal of Emerging and Selected Topics in Power Electronics, vol. 9, no. 2, pp. 1290-1301, April 2021, doi: 10.1109/JESTPE.2020.2969144.

\bibitem{Zhao}
P. Zhao, Z. Liu and J. Liu, "An Adaptive Discrete Piecewise Droop Control in DC Microgrids," in IEEE Transactions on Smart Grid, vol. 15, no. 2, pp. 1271-1288, March 2024, doi: 10.1109/TSG.2023.3302688.

\bibitem{Chandak2022}
S. Chandak and M. Rout, ``The implementation framework of a microgrid: A review,'' \textit{Int. J. Energy Res.}, vol. 46, no. 3, pp. 5053--5080, 2022.

\bibitem{Guerrero2011}
J. M. Guerrero, J. C. Vasquez, J. Matas, L. G. de Vicuña, and M. Castilla, ``Hierarchical control of droop-controlled AC and DC microgrids---A general approach toward standardization,'' \textit{IEEE Trans. Ind. Electron.}, vol. 58, no. 1, pp. 158--172, 2011.

\bibitem{rashwan2023}
A. Rashwan et al., ``Modified Droop Control for Microgrid Power-Sharing Stability Improvement,'' \textit{Sustainability}, vol. 15, no. 14, Art. no. 11220, 2023.

\bibitem{Lu2024}
M. Lu, M. Qin, W. Mu, J. Fang and S. M. Goetz, ``A hybrid gallium-nitride–silicon direct-injection universal power flow and quality control circuit with reduced magnetics,'' in \textit{IEEE Transactions on Industrial Electronics}, vol. 71, no. 11, pp. 14161--14174, Apr. 2024, doi: 10.1109/TIE.2024.3370999.

\bibitem{blaabjerg2006overview}
F. Blaabjerg et al., ``Overview of control and grid synchronization for distributed power generation systems,'' \textit{IEEE Trans. Ind. Electron.}, vol. 53, no. 5, pp. 1398--1409, 2006.

\bibitem{Middlebrook1976}
R. D. Middlebrook, ``Input filter considerations in design and application of switching regulators,'' \textit{IEEE IAS Annu. Meeting}, pp. 366--382, 1976.

\bibitem{kacetldc}
J. Kacetl et al., ``Design and Analysis of Modular Multilevel Reconfigurable Battery Converters for Variable Bus Voltage Powertrains,'' \textit{IEEE Transactions on Power Electronics}, vol. 38, no. 1, pp. 130--142, 2023. doi: 10.1109/TPEL.2022.3179285.

\bibitem{Kwasinski2011}
A. Kwasinski, ``Quantitative evaluation of DC microgrids availability: Effects of system architecture and converter topology design choices,'' \textit{IEEE Trans. Power Electron.}, vol. 26, no. 3, pp. 835--851, 2011.

\bibitem{Kwasinski2011b}
A. Kwasinski and C. N. Onwuchekwa, ``Dynamic Behavior and Stabilization of DC Microgrids with Instantaneous CPL,'' \textit{IEEE Transactions on Power Electronics}, vol. 26, no. 3, pp. 822--834, Mar. 2011, doi: 10.1109/TPEL.2010.2095869.

\bibitem{electric_springs}
S. Y. Hui, C. K. Lee, and F. F. Wu, ``Electric springs---A new smart grid technology,'' \textit{IEEE Trans. Smart Grid}, vol. 3, no. 3, pp. 1552--1561, 2012.

\bibitem{deng2021interaction}
W. Deng, Y. Pei, L. Kong, and Z. Zhao, ``Interaction and Stability Analysis of Multi-Converter Systems in DC Microgrids,'' \textit{IEEE Trans. Power Electron.}, vol. 36, no. 3, pp. 3291--3303, March 2021.

\bibitem{tu2023impact}
H. Tu, H. Yu, and S. Lukic, ``Impact of Virtual Inertia on DC Grid Stability with Constant Power Loads,'' \textit{IEEE Trans. Power Electron.}, vol. 38, no. 5, pp. 5693--5697, May 2023.

\bibitem{UltraCompactPowerBuffer}
M. Guacci et al., ``Ultra-compact power pulsation buffer for electric vehicle fast charging,'' \textit{IEEE Trans. Power Electron.}, vol. 36, no. 8, pp. 8963--8974, 2021.

\bibitem{SeriesVoltageCompensator}
Y. Wang et al., ``Series voltage compensator for differential power processing,'' \textit{IEEE Trans. Power Electron.}, vol. 36, no. 5, pp. 5851--5861, 2021.


\bibitem{overview_SOP}
W. Cao et al., ``Overview of soft open points for power distribution network: Concept, application and prospect,'' \textit{IEEE Trans. Power Syst.}, vol. 34, no. 3, pp. 1897--1908, 2019.

\bibitem{delta_SNOP}
J. Wang et al., ``Delta-type soft normally open point (S-NOP) for power distribution network,'' \textit{IEEE Trans. Power Del.}, vol. 35, no. 2, pp. 674--684, 2020.

\bibitem{series_shunt_SNOP}
Q. Qi et al., ``Series-shunt soft normally open point (S-NOP) for power distribution network,'' \textit{IEEE Trans. Power Del.}, vol. 36, no. 4, pp. 2143--2153, 2021.

\bibitem{hui2012electric}
S. Y. R. Hui, C. K. Lee, and F. F. Wu, ``Electric springs---A new smart grid technology,'' \textit{IEEE Transactions on Smart Grid}, vol. 3, no. 3, pp. 1552--1561, 2012.

\bibitem{lu2016optimal}
Y. Lu, J. Wu, N. Jenkins, and C. Wang, ``Optimal operation of soft open points in medium voltage electrical distribution networks with distributed generation,'' \textit{Applied Energy}, vol. 184, pp. 427--437, 2016.

\bibitem{chen2013transformerless}
Y. Chen, W. Wu, and B. Guo, ``A transformerless unified power flow controller based on the cascade multilevel inverter,'' \textit{IEEE Transactions on Power Electronics}, vol. 28, no. 5, pp. 2395--2403, 2013.

\bibitem{guan2015transformerless}
M. Guan, Z. Xu, and X. Liu, ``Transformerless unified power flow controller (TL-UPFC),'' \textit{IEEE Transactions on Power Electronics}, vol. 31, no. 2, pp. 1135--1147, 2015.

\bibitem{gyugyi1992unified}
L. Gyugyi, ``Unified power-flow control concept for flexible AC transmission systems,'' \textit{IEE Proc. C}, vol. 139, no. 4, pp. 323--331, 1992.

\bibitem{hingorani2000understanding}
N. G. Hingorani and L. Gyugyi, \textit{Understanding FACTS: Concepts and Technology}, IEEE Press, 2000.

\bibitem{zhang2012flexible}
X.-P. Zhang, C. Rehtanz, and B. Pal, \textit{Flexible AC Transmission Systems: Modelling and Control}, Springer, 2012.

\bibitem{peng}
H. Peng, J. Zhang, J. Zhou, G. Shi, J. Wang and X. Cai, ``Delta-Type Serial Shunt Soft Normally-Open Points With Wide Power Flow Regulation Range in Distributed Network,'' in \textit{IEEE Transactions on Industrial Electronics}, vol. 71, no. 7, pp. 6501--6511, July 2024, doi: 10.1109/TIE.2023.3310008.

\bibitem{Zhang}
J. Zhang et al., ``Series--Shunt Multiport Soft Normally Open Points,'' in \textit{IEEE Transactions on Industrial Electronics}, vol. 70, no. 11, pp. 10811--10821, Nov. 2023, doi: 10.1109/TIE.2022.3229375.

\bibitem{purgat2020power}
P. Purgat, \textit{et al.}, ``Power Flow Decoupling Controller for Triple Active Bridge Based on Fourier Decomposition of Transformer Currents,'' in \textit{Proc. of IEEE Energy Conversion Congress and Exposition}, 2020.

\bibitem{Lu2023}
M. Lu, M. Qin, J. Kacetl, E. Suresh, T. Long and S. M. Goetz, ``A Novel Direct-Injection Universal Power Flow and Quality Control Circuit,'' in \textit{IEEE Journal of Emerging and Selected Topics in Power Electronics}, vol. 11, no. 6, pp. 6028--6041, Dec. 2023, doi: 10.1109/JESTPE.2023.3321882.

\bibitem{Lu2024b}
M. Lu, W. Mu, M. Qin, J. Fang and S. M. Goetz, ``Grid impedance estimation and decoupling through a series–parallel direct-injection soft open point,'' in \textit{IEEE International Power Electronics and Motion Control Conference (IPEMC2024--ECCE Asia)}, vol. 10, pp. 4975--4979, May 2024, doi: 10.1109/IPEMC-ECCEAsia60879.2024.10567333.

\bibitem{Lu2023b}
M. Lu, Y. Fan, C. Zhang and S. M. Goetz, ``An Inductorless Direct-injection Power Control Circuit for the Distribution Grid,'' in \textit{IEEE Energy Conversion Congress and Exposition (ECCE)}, pp. 6279--6284, Dec. 2023, doi: 10.1109/ECCE53617.2023.10362491.

\bibitem{Liang2019}
B. Liang, L. Kang, J. He, F. Zheng, Y. Xia, Z. Zhang, Z. Zhang, G. Liu and Y. Zhao, ``Coordination control of hybrid AC/DC microgrid,'' in \textit{The Journal of Engineering}, vol. 2019, no. 16, pp. 3264--3269, Jan. 2019, doi: 10.1049/joe.2018.8505.

\bibitem{DPP}
J. Su and K. Li, ``Differential Power Processing Based Control Framework for Multiple Battery Energy Storage Systems in DC Microgrids,'' in \textit{IEEE Transactions on Sustainable Energy}, vol. 15, no. 4, pp. 2417--2427, Oct. 2024, doi: 10.1109/TSTE.2024.3421358.

\bibitem{Rivera2022}
S. Rivera et al., ``Partial-power converter topology for electric vehicle fast charging stations,'' \textit{IEEE Trans. Transport. Electrific.}, vol. 8, no. 2, pp. 1547--1559, 2022.

\bibitem{RiveraCharging}
S. Rivera et al. Charging infrastructure and grid integration for electromobility. Proceedings of the IEEE, vol. 111, no. 4, pp. 371-396, 2022.

\bibitem{MIL217}
U.S. Department of Defense, MIL-HDBK-217F: Reliability Prediction of Electronic Equipment, Washington, DC: U.S. Government Printing Office, 1989.

\bibitem{IEC 61709}
International Electrotechnical Commission, IEC 61709:2017 -- Electronic components reliability -- Reference conditions for failure rates and stress models for conversion. Geneva, Switzerland: IEC, 2017.

\bibitem{IEC 62380}
International Electrotechnical Commission, IEC 62380:2006 -- Reliability data handbook -- Universal model for reliability prediction of electronics components, PCBs, and equipment. Geneva, Switzerland: IEC, 2006.

\bibitem{CFC_DCCB}
A. Sajjadi, G. Kadkhodaei, M. Hamzeh, S. D. Tavakoli and O. Gomis-Bellmunt, ``An Integrated Multiport Circuit Breaker With Current Flow Controlling Capability for Meshed Multiterminal HVDC Grids,'' in \textit{IEEE Transactions on Power Electronics}, vol. 39, no. 8, pp. 9680--9693, Aug. 2024, doi: 10.1109/TPEL.2024.3396137.

\bibitem{voltage2010ch11}
\textit{Voltage-Sourced Converters in Power Systems}, Chapter 11: ``Static Compensator (STATCOM),'' Wiley, 2010, pp. 311--333, doi: 10.1002/9780470551578.ch11.


\bibitem{Khan2023}
D. Khan, M. Qais, I. Sami, P. Hu, K. Zhu, and A. Y. Abdelaziz, ``Optimal LCL-filter design for a single-phase grid-connected inverter using metaheuristic algorithms,'' \textit{Computers and Electrical Engineering}, vol. 110, 2023, Art. no. 108857, ISSN 0045-7906, doi: 10.1016/j.compeleceng.2023.108857.

\bibitem{Parker2014}
S. G. Parker, B. P. McGrath, and D. G. Holmes, ``Regions of Active Damping Control for LCL Filters,'' \textit{IEEE Transactions on Industry Applications}, vol. 50, no. 1, pp. 424--432, Jan.-Feb. 2014, doi: 10.1109/TIA.2013.2266892.

\bibitem{Muhlethaler2013}
J. Muhlethaler, M. Schweizer, R. Blattmann, J. W. Kolar, and A. Ecklebe, ``Optimal Design of LCL Harmonic Filters for Three-Phase PFC Rectifiers,'' \textit{IEEE Transactions on Power Electronics}, vol. 28, no. 7, pp. 3114--3125, July 2013, doi: 10.1109/TPEL.2012.2225641.

\bibitem{yazdani}
Amirnaser Yazdani; Reza Iravani, "Static Compensator (STATCOM)," in Voltage-Sourced Converters in Power Systems: Modeling, Control, and Applications , IEEE, 2010, pp.311-333, doi: 10.1002/9780470551578.ch11. 

\bibitem{Wang2016}
J. Wang, J. M. Guerrero, and F. Blaabjerg, \textit{Control Techniques for LCL-Type Grid-Connected Inverters}. Wiley-IEEE Press, 2016.



\bibitem{Ciobotaru2006}
M. Ciobotaru, R. Teodorescu, and F. Blaabjerg, ``A new single-phase PLL structure based on second order generalized integrator,'' \textit{IEEE Power Electronics Specialists Conference}, vol. 37, pp. 1--6, 2006.

\bibitem{Fang2019}
J. Fang, P. Lin, H. Li, Y. Yang and Y. Tang, ``An Improved Virtual Inertia Control for Three-Phase Voltage Source Converters Connected to a Weak Grid,'' in \textit{IEEE Transactions on Power Electronics}, vol. 34, no. 9, pp. 8660--8670, Sept. 2019, doi: 10.1109/TPEL.2018.2885513.

\bibitem{DroopControlEnhancement}
J. He et al., ``An enhanced droop control strategy for accurate power sharing in islanded microgrids,'' \textit{IEEE Trans. Power Electron.}, vol. 36, no. 7, pp. 7843--7854, 2021.

\bibitem{DroopControlLimitationsSolution}
M. C. Chandorkar et al., ``Control of parallel connected inverters in standalone AC supply systems,'' \textit{IEEE Trans. Ind. Appl.}, vol. 29, no. 1, pp. 136--143, 1993.


\bibitem{electric_springs_review}
Y. Yang, S.-C. Tan, and S. Y. R. Hui, ``A Review of AC and DC Electric Springs,'' \textit{IEEE Access}, vol. 9, pp. 14358--14371, 2021.




\bibitem{Lu2025}
M. Lu, W. Mu, M. Qin, A. Koehler, J. Fang and S. M. Goetz, ``Differential detection of feeder and mesh impedances through a series--parallel direct-injection soft open point,'' in \textit{IEEE Transactions on Power Electronics}, vol. 49, no. 1, pp. 1964--1973, Jan. 2025, doi: 10.1109/TPEL.2024.3456071.


\bibitem{Kundur1994}
P. Kundur, \textit{Power System Stability and Control}, McGraw-Hill, 1994.

\bibitem{Saleh2017}
K. A. Saleh, A. Hooshyar, and E. F. El-Saadany, \textit{Hybrid Passive-Overcurrent Relay for Detection of Faults in Low-Voltage DC Grids}, \textit{IEEE Transactions on Smart Grid}, vol. 8, no. 3, pp. 1129--1137, 2017. DOI: 10.1109/TSG.2015.2477482.

\end{thebibliography}
\end{document}